# CHOKeD: A Fair Active Queue Management System

Thesis submitted in partial fulfillment for the degree of

**Master of Science in Computer System Engineering**

**By**

**SANAULLAH MANZOOR**

**CS1301**

**FACULTY OF COMPUTER SCIENCE AND ENGINEERING**

**GHULAM ISHAQ KHAN INSTITUTE OF ENGINEERING SCIENCES AND TECHNOLOGY**

# CERTIFICATE OF APPROVAL

## CERTIFICATE OF APPROVAL

It is certified that the research work presented in this thesis titled "CHOKeD: A Fair Active Queue Management System" was conducted by Mr. Sanaullah Manzoor under my supervision and that in my opinion, is fully adequate, in scope and quality, for the degree of MS in Computer System Engineering.

(Supervisor)
Dr. Masroor Hussian
Faculty of Computer Sciences & Engineering, Ghulam Ishaq Khan Institute of Engineering Scieneces & Technology,
Topi-23460, Swabi, Pakistan.

(Co-Supervisor)
Dr. Ghulam Abbas
Faculty of Computer Sciences & Engineering, Ghulam Ishaq Khan Institute of Engineering Scieneces & Technology,
Topi-23460, Swabi, Pakistan.

(Internal Examiner)
Dr. Ahmar Rashid
Faculty of Computer Sciences & Engineering, Ghulam Ishaq Khan Institute of Engineering Scieneces & Technology,
Topi-23460, Swabi, Pakistan.

(Dean)
Prof. Dr. Khalid J. Siddiqui
Faculty of Computer Sciences & Engineering, Ghulam Ishaq Khan Institute of Engineering Scieneces & Technology,
Topi-23460, Swabi, Pakistan.

(External Examiner)
Dr. Sajjad A. Madani
COMSATS, Institute of Information Technology,
Islamabad, Pakistan.

Dedicated

To

My beloved parents, family members, respected teachers and friends

# ABSTRACT

Fairness is the significant factor to sustain best effort delivery of network services. Now-a-days, real-time multimedia applications have evolved largely over the Internet. Most of multimedia services are unresponsive during network congestion. Unresponsive traffic streams steal most network bandwidth and starve out some other flows which are responsive during network congestion. In the presence of these unresponsive traffic flows, protection of responsive flows has become a major issue. Many Active Queue Management (AQM) based solutions have been recommended to protect responsive traffic flows from unresponsive ones and to ensure fairness among all traffic flows.

The thesis proposes a novel AQM scheme CHOKeD, to deliver fairness among all flows of a congested link. It is a completely stateless approach. CHOKeD is based on dynamic drawing factor to penalize unresponsive traffic. It successfully protects responsive flows in the presence of unresponsive flows. CHOKeD features such as fairness, high throughput of responsive traffic and stateless design, are encouraging factors for its deployment over the edge as well as the network core routers. Extensive simulations have been carried out to evaluate its performance under real-time network scenarios.

# ACKNOWLEDGMENT

First of all, thanks to ALLAH ALMIGHTY, for His countless blessings upon me and numerous Darood (payer) upon last messenger of ALLAH, Prophet MUHAMMAD (PBUH). Additionally, I would like to pay cordial gratitude to my supervisor Dr. Masroor Hussain for his vision, guidance and encouragement throughout my work. I am very thankful to my co-advisor Dr. Ghulam Abbas for his kind support and sharing valuable time during all the phases of research and guidance for the successful completion of the thesis. I am also obliged to Dean FCSE, Dr. Khalid J. Siddiqui for his valuable suggestions and guidance during thesis writing.

I am greatly thankful to my loving parents and siblings for their unconditional love, support and trust. I would like to acknowledge GIK Institute and Faculty of Computer Science and Engineering, on awarding MS fellowship and scholarship for my MS studies, as well as providing a conductive environment for study and research. Finally, I would like to thank all the teachers, fellow colleagues and graduate students, especially Faizan-e-Mustafa and Ahsan Saeed, who have supported and assisted me throughout my graduate years at the GIK Institute. Special thanks to my elder brothers Dr. Tareq Manzoor, Tahir Manzoor and younger brother Habibullah Manzoor for their suggestions and technical support.

# TABLE OF CONTENTS

# LIST OF FIGURES

# LIST OF TABLES

## LIST OF ABBREVIATIONS AND SYMBOLS

| ABBREVIATION | TERM |
|---|---|
| AQM | Active Queue Management |
| PQM | Passive Queue Management |
| QoS | Quality of Service |
| ECN | Explicit Congestion Notification |
| RED | Random Early Detection |
| TCP | Transport Control Protocol |
| UDP | User Datagram Protocol |
| BW | Bandwidth |
| FIFO | First In First Out |
| EWMA | Exponential Weighted Moving Average |
| NS-2 | Network Simulator 2 |
| Tcl | Tool Command Language |

| SYMBOL | DETAIL |
|---|---|
| $Q_a$ | Average queue size |
| $Q_c$ | Current queue size |
| $T_{min}$ | Minimum queue threshold |
| $T_{max}$ | Maximum queue threshold |
| $w_q$ | RED queue weight |
| $B$ | Buffer size |
| $w_n$ | TCP expected congestion window size in packets |
| $C_p$ | Link Capacity |
| $R_t$ | Round trip time |
| $P_d$ | Dropping probability |

# CHAPTER 1

## 1 INTRODUCTION

The majority of the traffic in the Internet today is carried by the Transmission Control Protocol (TCP). TCP is the protocol of choice for the widely used World Wide Web (HTTP), file transfer (FTP), TELNET, and email (SMTP) applications, due to its reliability. In practice, TCP has been widely deployed because of its congestion avoidance algorithm, introduced by Jacobson (Jacobson, 1988). This mechanism helps the traffic sources to determine the available network bandwidth and regulate their transmission rates accordingly. The basic idea of this approach is to have the source gradually increase its transmission rate until the available bandwidth and reduce its transmission rate when congestion is detected. TCP sources increase or decrease their rates by use of a variable called congestion window, denoted as $C_w$. TCP increases the size of $C_w$ linearly and decrease it in a multiplicative manner, if congestion is detected. TCP's this congestion control strategy keeps the network from being overloaded and congestion collapse. We can call TCP driven flows "responsive" as they lower their rate on account of network congestion.

These days, a number of new applications such as multimedia (audio/video streaming applications, voice over IP, etc.) applications are being widely deployed over the Internet. These applications are supported by the User Datagram Protocol (UDP) rather than TCP. UDP lacks end-to-end flow control as well as congestion control mechanism. Under UDP, sources independently adjust their transmission rates and do not take account of the network congestion. Due to the increased deployment trend of UDP and its inability to control congestion, these applications result in two major problems for the Internet, congestion collapse and unfairness (Floyd *et al.*, 1999).

Congestion collapse occurs when senders keep sending their packets without taking into account of the network congestion state. These packets are discarded before reaching destinations by bottleneck link. This is a serious problem because a large amount of bandwidth is wasted due to these undelivered packets (Floyd *et al.,* 1999). Such UDP driven sources are termed as “unresponsive” flows. Unresponsive because these flows are ignorant of network congestion and fail to adjust transmission rates accordingly.

## 1.1 PROBLEM OF UNFAIRNESS

The unfairness problem emerges when UDP driven traffic, *i.e.,* unresponsive flows steal a large amount of bandwidth of the bottleneck link as compared to the well-behaving responsive TCP flows. Because after detecting congestion, TCP sources start reducing their transmission rates while UDP sources keep sending at initial rate and thus, occupy a portion of bandwidth left by responsive TCPs. This behavior continues and under persistent congestion, eventually, unresponsive flows may successfully consume the entire bandwidth. Although some streaming applications claim some level of responsiveness to control network congestion, but these are still more aggressive than well-behaving TCP flows and cause unfairness (Chatranon *et al.,* 2004).

To handle this unfairness problem, router-based fairness control solutions have been proposed. Router based algorithms are deployed inside the network router to regulate flows that cause unfairness. To maintain network fairness, these algorithms introduce some punishment mechanisms for unresponsive flows. These punishment mechanisms help the responsive flows to attain a better bandwidth share and lower the bandwidth of unresponsive flows.

## 1.2 ROUTER BASED UNFAIRNESS CONTROL

Generally speaking, there are two kinds of router based AQM mechanisms to enforce fairness, first one is per-flow fair queuing or scheduling and second type is router based FIFO queue management. The former approach is stateful and maintains a number of queues for all flows and is complex as well (Nagle, 1987), while the latter approach maintains only single FIFO queue for all flows and may be completely stateless approach. Well-known per-flow queuing schemes for router fairness are Start-time Fair Queuing (Goyal *et al.*, 1996), Self-Clocked Fair Queuing (SCFQ) (Golestani, 1994). These techniques maintain states for each flow and have to identify each flow. Also, these schemes possibly manage large number of queues.

In this thesis, active queue management based approach is adopted to solve the unfairness problem.

### 1.2.1 QUEUE MANAGEMENT

Generally, to maintain a router's buffer or the queue, there are two types of queue management schemes for the routers, first one is Active Queue Management (AQM) like Random Early Detection (RED) (Floyd and Jacobson, 1993) and other one is Passive Queue Management (PQM) like Drop Tail (Braden *et al.,* 1998). An old PQM based technique was Drop Tail, which maintains the queue of maximum length in terms of packets. In PQM, it accepts packets, until the maximum length of the queue is reached and after that it drops all arriving packets. As most recently arrived packet at the tail has to be dropped, that is why this PQM scheme is called Drop Tail. In earlier days, Drop Tail adopted widely for network routers, but later some serious issues of it were revealed. There are two major drawbacks of Drop Tail, one is lockout and the other one is full queue. In lockout, there may be certain situations in which one or more flows can monopolize all queue space and restricting other flows to get space in the queue. The second drawback of Drop Tail, *i.e.*, full queue causes global synchronization, such that when the queue is full, packets arrive at the router in bursts more often, and all busty packets will be dropped. In other words, Drop Tail is bias to busty traffic (Braden *et al.*, 1998).

### 1.2.2 ACTIVE QUEUE MANAGEMENT

Fairness among all flow of network can be done through AQM (Braden *et al.*, 1998). One of the widely adopted and successful techniques for router's active queue management is Random Early Detection (RED), recommended by Internet Research Task Force (IRTF). AQM like RED is completely stateless, simple and maintains only one FIFO queue. RED solves full queue and lockout, drawbacks of Drop Tail (Braden *et al.*, 1998). RED calculates exponentially weighted moving average queue size. In RED queue length is marked with two thresholds: minimum threshold and maximum threshold. If the average queue size is less than the minimum threshold, RED does not mark or drop any arrived packet. But if the average queue size is greater than maximum threshold, RED drops all arrived packets. If the average queue size is between maximum and minimum thresholds, RED marks arriving packets with certain probability. This probability helps to control the average queue size through marking of packets (Floyd and Jacobson, 1993).

In RED, all flow share single FIFO queue and during congestion, it probabilistically drops packets. This dropping probability increase as congestion of FIFO queue increases. Small average size reduces queue delays. RED successfully enables fairness among responsive TCP flows, when congestion occurs, but fails to punish or restrict unresponsive flows (Pan *et al.*, 2000).

## 1.3 PROBLEM STATEMENT

Some active queue management based solutions like RED completely fail to control unfairness in the presence of unresponsive or malicious flows. Unresponsive flows occupy more percentage of bandwidth and starve out responsive flows, for same packet drop rate. These unresponsive flows ruin the concept of Internet bandwidth fairness and cause congestion of shared link. To overcome such limitations and compel improvement in traditional AQM techniques, there is a need to develop a fair active queue management mechanism.

## 1.4 SCOPE OF THESIS

The main features and the contributions of this thesis are described as follow:

- A new router based AQM scheme is proposed to protect responsive flows and ensure fairness among all flows.
- This thesis focuses on simplicity and scalability of AQM protocol to control unfairness. Therefore, a stateless approach is adopted.
- Simulations are performed to validate the proposed protocol CHOKeD against other AQM protocols namely RED (Floyd and Jacobson, 1993), CHOKe (Pan *et al.*, 2000) and gCHOKe (Eshete and Jiang, 2013).

Fig. 1.1 graphically outlines the summary of the subject area of this thesis in the grey color.

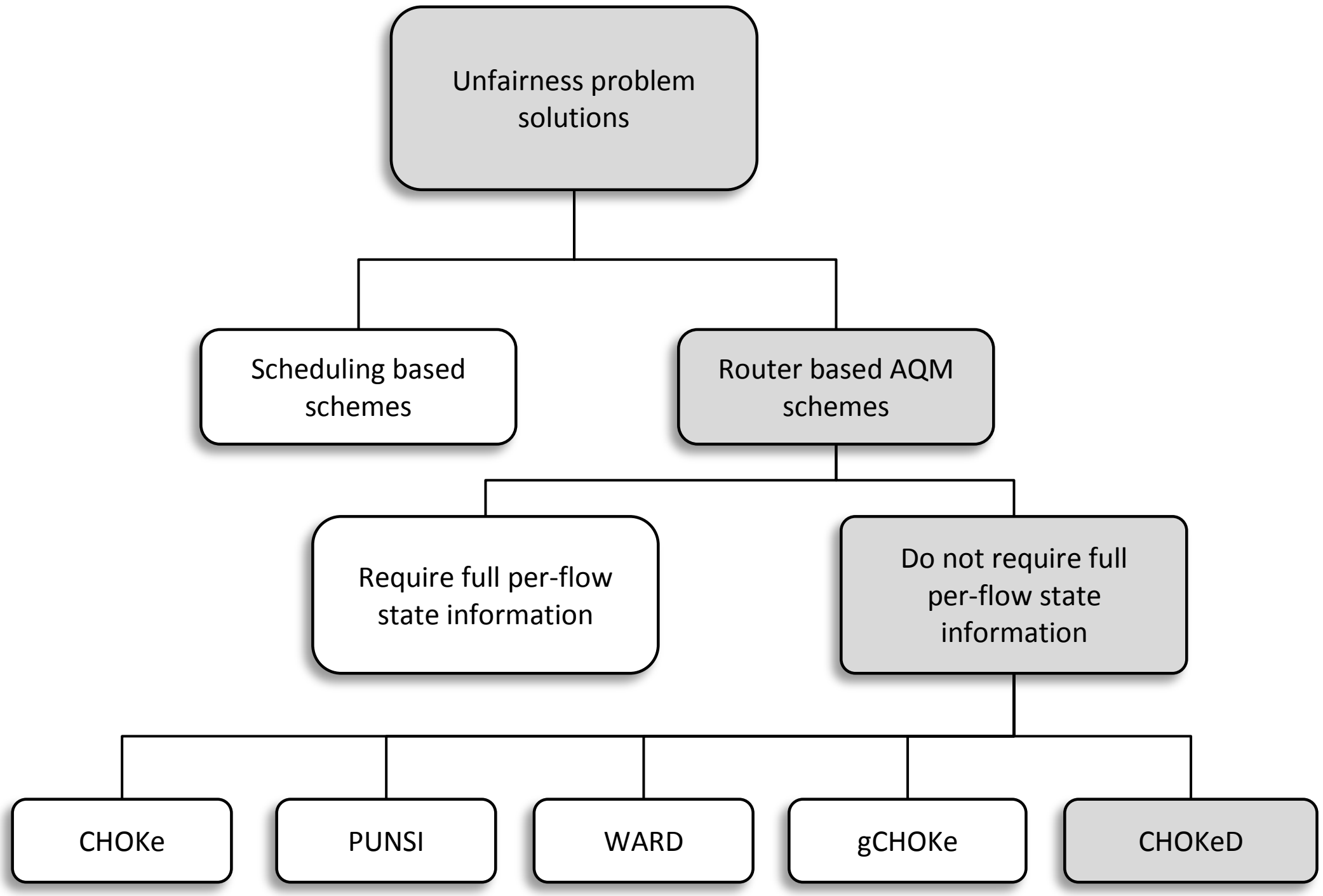

FIG. 1.1 SCOPE OF THE THESIS

## 1.5 THESIS ORGANIZATION

The rest of the thesis is organized as follow:

- Chapter 2 presents literature review, portraying stateless active queue management schemes.
- Chapter 3 covers the proposed scheme about fairness in gateway routers.
- Chapter 4 is about simulations and results.
- Finally, Chapter 5 concludes the thesis and presents some future directions.

# CHAPTER 2

# 2 LITERATURE REVIEW

This thesis covers fundamentals of TCP congestion control, in the background section, firstly. Secondly, it discusses traditional and current fair active queue management based schemes to control unfairness problem due to unresponsive flows at network router. Many algorithms developed that claim fairness among currently residing flows of underlying router. These algorithms focus on the protection of friendly flows from un-friendly, *i.e.*, unresponsive flows to establish a certain fairness level in network.

## 2.1 BACKGROUND

TCP provides ordered, reliable and connection-oriented delivery of services. TCP congestion and flow control mechanism vary transmission rate according to available bandwidth of network. Sender maintains a state variable called congestion window ($C_w$). Congestion window determines the amount of data a sender can send at a given time, according to network congestion level. On the other hand, receiver has Advertised Window that reflect receiver's amount of data capacity. A sender maximally can send the minimum of both congestion window and advertised window size, data capacity at a given time.

Initially, TCP sources are unaware of network congestion level until the Round Trip Time (RTT) out or packet drop. Therefore, TCP sources slowly increase transmission rate exponentially by increasing congestion window size until congestion occurs or a Slow Start Threshold is (ssthresh) reached. This TCP state is known as the slow start phase, in which TCP source doubles its congestion window size after every round trip time. Once

TCP attains ssthresh, now it enters in Congestion Avoidance phase. In the congestion avoidance phase, sender increases congestion window by $\frac{1}{C_w}$ after receiving acknowledgement. This is known as Additive Increase and Multiplicative Decrease (AIMD). Fig. 2.1 shows TCP AIMD, slow start and congestion avoidance phases. During the congestion avoidance phase if congestion occurs, TCP source cut down ssthresh to half of the current congestion window size $C_w$, and sets $C_w$ equals to 1 packet size, here it enters into the slow start phase again. Currently, there are many well-known TCP versions are available such as Tahoe, Reno, SACK, Newreno, and Vegas (Jacobson, 1990).

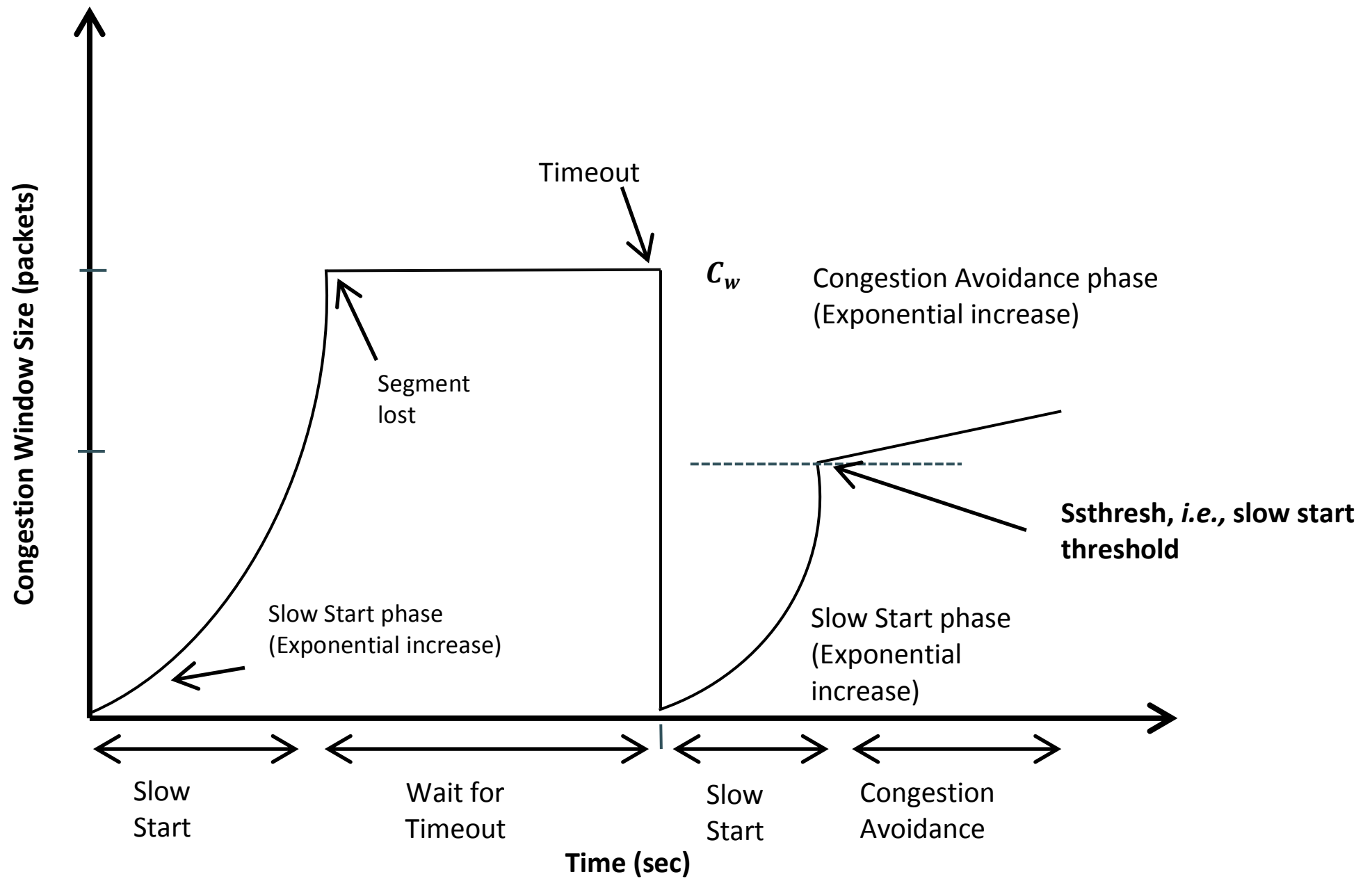

FIG. 2.1 TCP ADDITIVE INCREASE AND MULTIPLICATIVE DECREASE

Due to TCP congestion control mechanism, all TCP sources try to gain fair network bottleneck link bandwidth. TCP driven flows behave well-mannered, if congestion is detected and adjust their sending rate accordingly. This mechanism leads to the responsiveness of a flow to be fair under bottleneck link and avoid unfairness. UDP driven all flows lack congestion control mechanism, hence cause unfairness among all network flows. UDP's this behavior is categorized as unfriendly or unresponsive. Later sections cover router based AQM solutions for unfairness problem.

## 2.2 ACTIVE MANAGEMENT SCHEMES

During the last two decades, many router based solutions proposed to solve unfairness problem of Random Early Detection (Floyd and Jacobson, 1993). These schemes are based on active queue management principle. These schemes can be classified as (1) stateless, (2) partially stateful and (3) completely stateful. Partially sateful and completely stateful schemes are complex and has issues about their implementation on core as well as backbone routers as they require full or partial per-flow state information (Chatranon *et al.,* 2004). Type (2) and (3) schemes are complex in terms of space and time complexity. Stateless protocols ensure fairness without maintaining any states of residing flows. These are simple and scalable approaches and can be deployed on the edge as well as core (backbone) routers (Adams, 2013). The literature survey in this thesis only focuses on stateless techniques that can enforce fairness among all flows on a bottleneck link.

### 2.2.1 CHOKE

**Pan** *et al.,* (2000) proposed a technique called CHOKe (CHOose and Keep for responsive flows, CHOose and Kill for unresponsive flows) to solve RED problems. CHOKe does not keep records or states of unresponsive flow, and it is a completely stateless. Working mechanism of CHOKe is given as fellow: it calculates exponential weighted moving average based average queue length as in RED. CHOKe also defines same thresholds as in RED, minimum threshold $T_{min}$ and maximum threshold $T_{max}$ . No packet is dropped if the average queue size $Q_a$ is less than $T_{min}$ and if the average queue size is greater than $T_{max}$ all packets are dropped. If $T_{min} < Q_a < T_{max}$, CHOKe draws a packet from the queue randomly and compare it with an arriving packet. If both having same flow id, it drops both (arrived packet and packet from queue). If flow id's of both does not match, arrived packet is dropped with RED's defined probability (Pan *et al.*, 2000).

### 2.2.3 SCHOKe

**Yan** and **Guangzhao** (2001) proposed SCHOKe that uses CHOKe and Stabilized Random Early Detection (SRED) (Ott *et al.*, 1999), approach for the punishment of unresponsive flows. SCHOKe is also stateless and use a single FIFO buffer. SCHOKe first calculates hit probability "p_hit (t)" when a packet arrives at buffer, this probability is same as of SRED. When arrived packet, and randomly packet chosen from queue matched as having same flow id, hit comes. SCHOKe's mechanism is a little bit different from the SRED as it compares more than one packet (k packets) from the queue. If these k packets and arrived packet both belong to the same flow, that is also a hit. Where, k is independent of SRED's instantaneous buffer occupancy. There are no minimum and maximum thresholds like RED in SCHOKe. Dropping probability is updated by value of p_hit (t). If B is queue limit and $Q_c$ is buffer occupancy, then packet dropping or admitting criteria is as: if $Q_c$ is less than $\frac{1}{6} * B$ then admit all arrived packets. If $Q_c > \frac{1}{6} * B$, SCHOKe selects k packets from the queue. These k packets and arrived packets cause a hit, it can drop all these else if no hit occurs arrived packet is dropped with previous drop probability (Yan and Guangzhao, 2001).

### 2.2.4 xCHOKe

**Chhabra** *et al.*, (2002) proposed xCHOKe that is a modification of original CHOKe and its working mechanism is described as: Average queue length is calculated on arrival of each new packet. If the average queue size is less than $T_{min}$ no packet is dropped, all got placed in the buffer. If the average queue size becomes greater than $T_{max}$ xCHOKe draws a packet from the FIFO queue randomly and compare it with arrived packet. If both packets are from same flow stream, both are dropped and that is a "hit". If these do not match, packet from the buffer is the place back and arrived packet admitted with provability based on level of congestion. xCHOKe is different from original CHOKe as it saves "hits" into a table called "lookup table". On arrival of each incoming packet xCHOKe scans "lookup table" and if it found same flow id in table and same of arriving packet. Update table as "table hit" and drop arrived packet. If the arriving packet and packet from the queue have different flow ids, xCHOKe creates a new row for this flow id (Chhabra *et al.*, 2002).

### 2.2.5 Self-Adjustable Choke (SAC)

**Jiang** *et al.*, (2003) proposed a router based AQM solution for the unfairness problem, named as Self-Adjustable CHOKe (SAC). Jiang *et al.*, (2003) revealed that CHOKe although penalize unresponsive flows, but still it is not fair enough. They describe two shortcoming of CHOKe: one is unfairness among UDP flows and another one, the number of selected candidates for newly arrived packet's comparison. SAC working mechanism is given as: it divides region between 0 and $T_{max}$ into k number of regions. If queue size is in between maximum and minimum threshold, say in region named $'i'$, then $i$ number of candidates will be selected for comparison and will be dropped if matched. The value of k is described by two parameters P and R. R is the probability that arriving packet is UDP. And P is the probability that arriving packet and randomly packet selected from the queue has same flow id. UDP and TCP flows are treated differently in case of SAC, for TCP on congestion notification; SAC behaves like original CHOKe (Jiang *et al.,* 2003).

### 2.2.6 ECHOKE

**Xu** *et al.,* (2004) proposed ECHOKe that is an extension of CHOKe but instead of RED it uses dropping probability criteria of Random Exponential Marking (REM) (Athuraliya *et al.,* 2001). ECHOKe also defines same average queue size thresholds as in RED, $T_{min}$ and $T_{max}$. No packet is dropped if the average queue size $Q_a$ is less than $T_{min}$ and if the average queue size is greater than $T_{max}$ all packets are dropped. If $T_{min} < Q_a < T_{max}$, ECHOKe draws a packet from queue randomly and compare it with arriving packet. If both having same flow id, it drops both (arrived packet and packet from queue). Simply in other words, rest of all working is same as in original CHOKe except dropping probability (Xu *et al.,* 2004).

### 2.2.7 CHOKEW

**Wen** *et* al. (2005) proposed an AQM scheme called as CHOKeW, to protect TCP flows from unresponsive UDP flows. CHOKeW is stateless and has the ability to work in core routers. CHOKeW defines a term “drawing factor”. Drawing factor is a variable that determines when to pick a packet (or packets) and how many packets have to draw from the buffer for comparison. Drawing factor controls time as well as the maximum number of packets to be drawn. Let us have $i$ active flows in the queue then drawing factor $p_i$, where, is $p_i > 0$. $p_i$ is maximum randomly draws from the buffer on arrival of the packet from flow $i$. CHOKeW also controls the priority of flows by introducing factor $w_i(w_i \geq$ 1). Drawing factor for low priority traffic is $p_0$. A flow with higher priority has less chance that its packets to be victim of dropping than a flow with less priority. When value of $p_i$ is such that $0 \leq p_i < 1$, it draws one packet from the buffer. When $p_i \geq 1$, then one or more than one candidate packets to be drawn from the queue (Wen *et al.*, 2005).

### 2.2.8 PENALIZING UNRESPONSIVE FLOWS (PUNSI)

**Yamaguchi** and **Takahashi** (2007) proposed PUNSI (Penalizing Unresponsive flows with Stateless Information) to minimize unresponsive flows dominance over responsive. PUNSI penalizes packets with higher probability from unresponsive flows than those from responsive ones Unresponsive flows have higher rate traffic usually than responsive flows. Without recording per flow information, PUNSI provides a fair share among active flows at the router. PUNSI highlights two drawbacks of CHOKe: CHOKe not only punish or penalize higher bandwidth unresponsive flow, but also penalize responsive flows like TCP. Secondly, CHOKe performs poorly when there are fewer packets of unresponsive flows in queue (Yamaguchi and Takahashi, 2007).

### 2.2.9 WARD

**Ho** *et al.,* (2007) proposed a router based AQM scheme called as WARD. WARD allocates every position in FIFO queue a weighted value that can be maximally of value 1. When a packet arrives at the buffer, WARD compares arrived packet's possible position with a random number that can also be not greater than 1. If the random number is greater than the weighted value of position, then arrived packet is admitted. Otherwise, WARD takes two packets from the queue randomly and compare with the arrived packet. If all packets belong to same flow, it drops all packets. If arriving packet's flow id is not same as that of randomly drawn packets but these two packets belongs to same flow, it drops both packets. If none of three matched, both packets from the queue are retained and arrived packet is also admitted. Lastly, if the buffer is full newly arrived packet is discarded (Ho *et al.*, 2007).

### 2.2.10 RECHOKe

**Govindaswamy** *et al.,* (2007) proposed RECHOKe that is an extention of CHOKe, RED-PD (Mahajan *et al.*, 2001) and xCHOKe (Chhabra *et al.,* 2002). RECHOKe working is given as: when a packet arrives at RECHOKe queue, if the average queue size is less than the minimum threshold, all arriving packets accepted. If the average queue sizes larger than the maximum threshold value then all arriving packets are dropped. If the average queue size is in between maximum and minimum threshold value it uses the concept of xCHOKe, RECHOKe searches flow id of arrived packet in the lookup table. If the same id is present in the table, arrived packet is dropped and counter is incremented. If arrived packet and a packet from queue have same ids, then this flow's counter is set to one. If flow is already in the lookup table, then despite of setting for one, its value is incremented. RECHOKe is different from here with xCHOKe that RECHOKe let the arrived packet to be queued rather than drop (Govindaswamy *et al.,* 2007).

### 2.2.11 Hierarchical Rate Code

**Wang** (2010) proposed a fairness mechanism based on CHOKe "A fair active queue management algorithm based on hierarchical rate code". Its working mechanism is given as: On arrival of a packet at the queue, it measures flow rates to detect network congestion. According to Wang (2010), congestion is occurred, if flow rate is greater than the output flow rate and current queue size is greater than half of the queue buffer size. If network is congested, it takes a packet randomly from the queue and starts match drop comparisons. If packet from the queue and arrived packet both have same flow ids, it drops both packets else, only arrived packet is dropped. If there is no congestion detected on arrival of the packet, it admits new packet (Wang, 2010).

### 2.2.12 GEOMANTIC CHOKE (GCHOKE)

**Eshete** and **Jiang** (2013) proposed another original CHOKe extension name as geometric CHOKe. When an incoming packet arrives at router: A packet form FIFO queue is selected randomly. If an incoming packet and the selected packet belong to the same flow id, gCHOKe selects new packet form queue randomly, if match again selects another one and so on. Only two conditions can stop the packet matching process. Firstly, when incoming packet, and the selected packet both belong to a different flow type. Secondly, when the maximum number of "*maxcomp*" trial has been executed. All matched the packet, and selected packets are dropped. In case of miss-match of arrived packet and packet from the queue, the packet is marked with probability same as described in RED and its admission depends upon this probability (Eshete and Jiang, 2013).

## 2.3 MOTIVATION

Plenty of research has been done in the area of AQM as a solution for the unfairness problem. The main objective of this research is to protect responsive flows from unresponsive flows and allocate fair bandwidth among all flows. Router based AQM solutions must have to be simple in terms of memory and time complexity, because, only stateless AQM systems are encouraged to deploy over the network routers (Adams, 2013). CHOKe (Pan *et al.,* 2000) suggests simple and stateless solution for the unfairness problem. CHOKe gives an innovative idea of drawing single candidate packet form the queue to compare with incoming packets. CHOKe outperforms as compared with traditional AQM schemes like RED. gCHOKe's methodology improves CHOKe's performance without adding any record for number of flows residing in the queue. In terms of fairness and high throughput of responsive flows, there is a need for improvement in both CHOKe and gCHOKe's candidate packet selection criteria to add more effectiveness in the presence of unresponsive flows.

This thesis is motivated by CHOKe to punish unresponsive flows and ensure fairness with minimal memory and time complexity.

# CHAPTER 3

# 3 CHOKE DESCENDANT (CHOKED)

This chapter presents a novel active queue management based fairness protocol called CHOKeD. The main objective of this protocol is to ensure fairness among all flows of a network router. This thesis also focuses on the scalability of the protocol with router buffer size. Practically vendors design different buffer sized based routers depending upon the requirements. Thus, during designing of this AQM technique, fairness among all flows and the size of a router's buffer is taken into account.

## 3.1 INTRODUCTION

CHOKeD is introduced to improve the performance of CHOKe (Pan *et al.,* 2000) and gCHOKe's (Eshete and Jiang, 2013) and as a solution to their shortcomings. Although CHOKe and gCHOKe punish unresponsive flows, there is still room for improvements in terms of fairness and protection of responsive flows.

Before discussion of CHOKe and gCHOKe's shortcomings, let's look into the network router's buffer during packets from different flows, are arriving and leaving single FIFO queue. Router's buffer diagram is given in Fig 3.1.

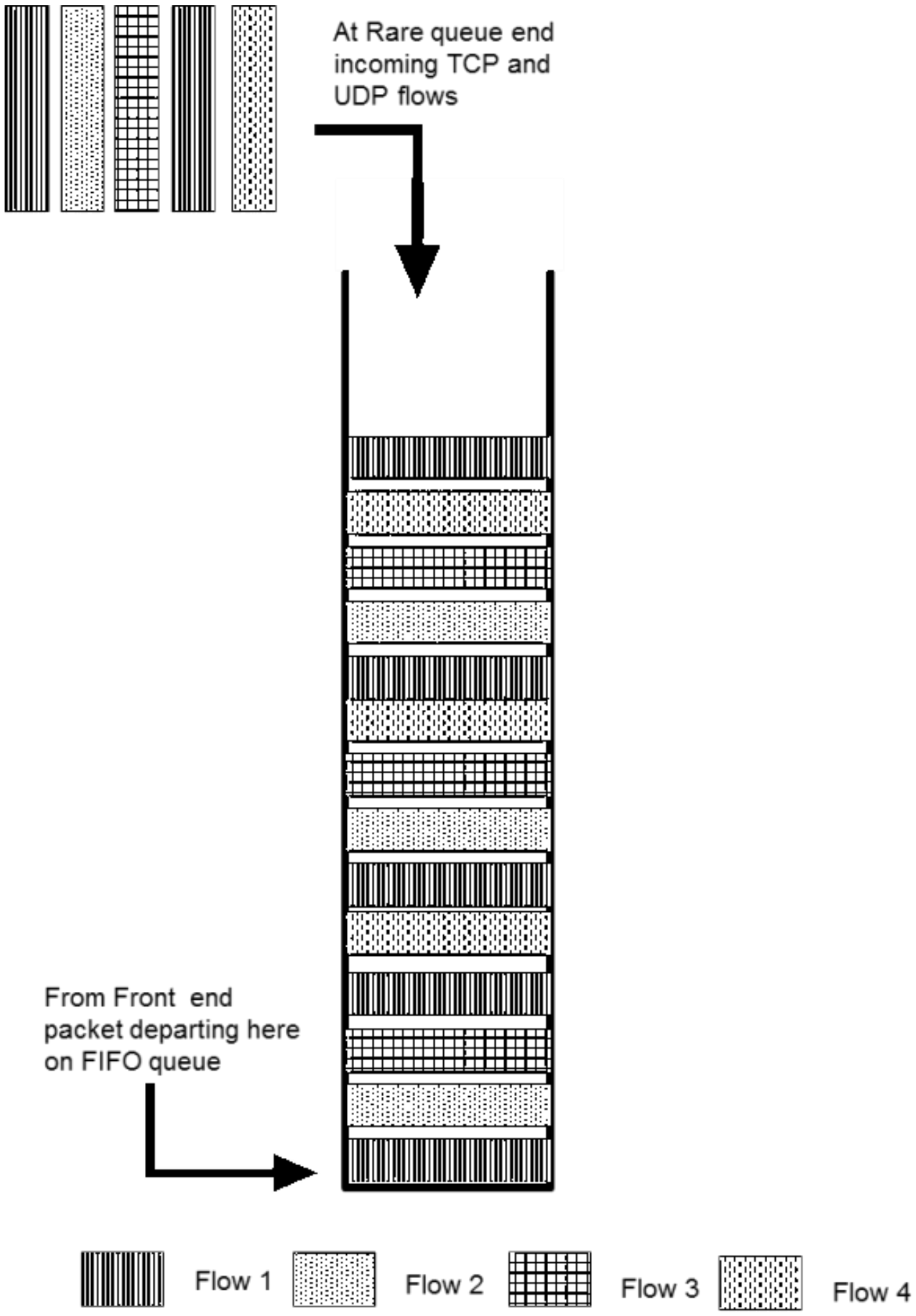

FIG. 3.1 PACKETS ARRIVING AND LEAVING THE BUFFER

Fig 3.1 describes the different types of flows arriving at the rear end of the queue. Flow 1, Flow 2, Flow 3 and Flow 4, represent different type of traffic flows in the queue. Different pattern schemes represent different flows based on their flow ids.

The following section confers some serious shortcomings of CHOKe and gCHOKe.

### 3.1.1 CHOKE'S SHORTCOMINGS

While extensive studying about CHOKe, these following shortcomings are observed:

1. One drawback of CHOKe is that it does not show concern with the location of arriving packet to the queue, and the position of drawing packet form queue. Experimental results reveal that unresponsive packets are more accumulated at the rear queue end than the front end. CHOKe lacks this, candidate packet selection explicitly from the region where unresponsive flows are more in number. To verify this, the experiment is performed, with 33 responsive TCP flows and 1 unresponsive UDP. NS2 simulation is carried out, using network topology shown in Fig. 3.2, with original CHOKe and with modification in CHOKe. This topology contains sender nodes at the left side, receiver nodes at the right and bottleneck link in the middle.

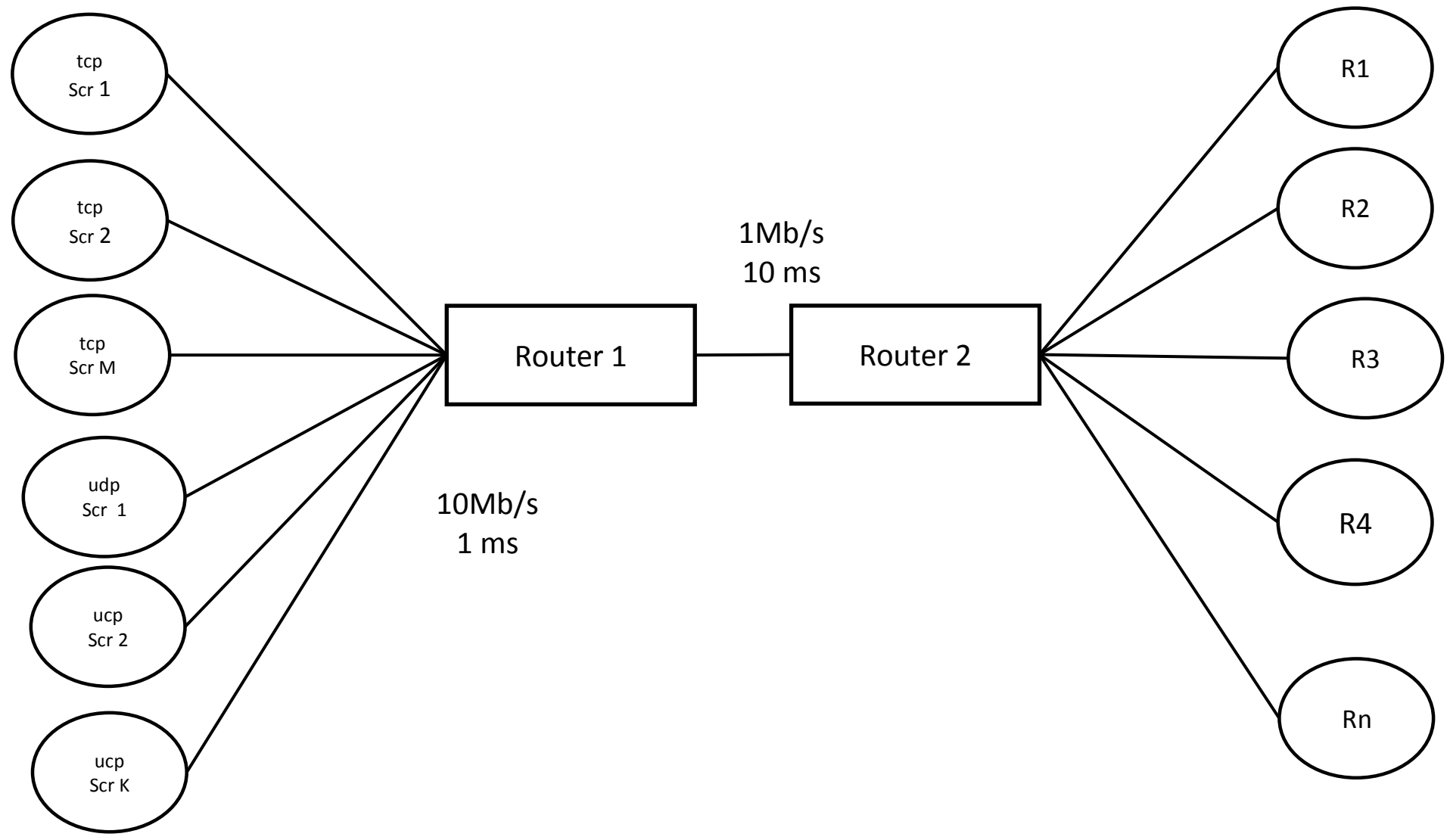

FIG. 3.2 SIMULATION TOPOLOGY

Senders have TCP and UDP based traffic. UDP is driven by Constant Bit Rate (CBR), represents the unresponsive traffic and TCP are driven by FTP applications. The modified version draws one packet from the rear queue region only, to compare with arriving packet. Fig. 3.3 and 3.4 shows the results of this simulation. From Fig. 3.3, UDP throughput under CHOKe is 0.127 Mb/s. From Fig 3.4, under CHOKe modified version, UDP throughput is 0.112 Mb/s. While ideal fair share is 0.028 Mb/s. Modified version's throughput is closer to ideal share. Decrease in UDP throughput under modified version is due to more successive match-drop than under original CHOKe. Thus, the more number of UDPs packets are accumulated at the rear end than at the front end.

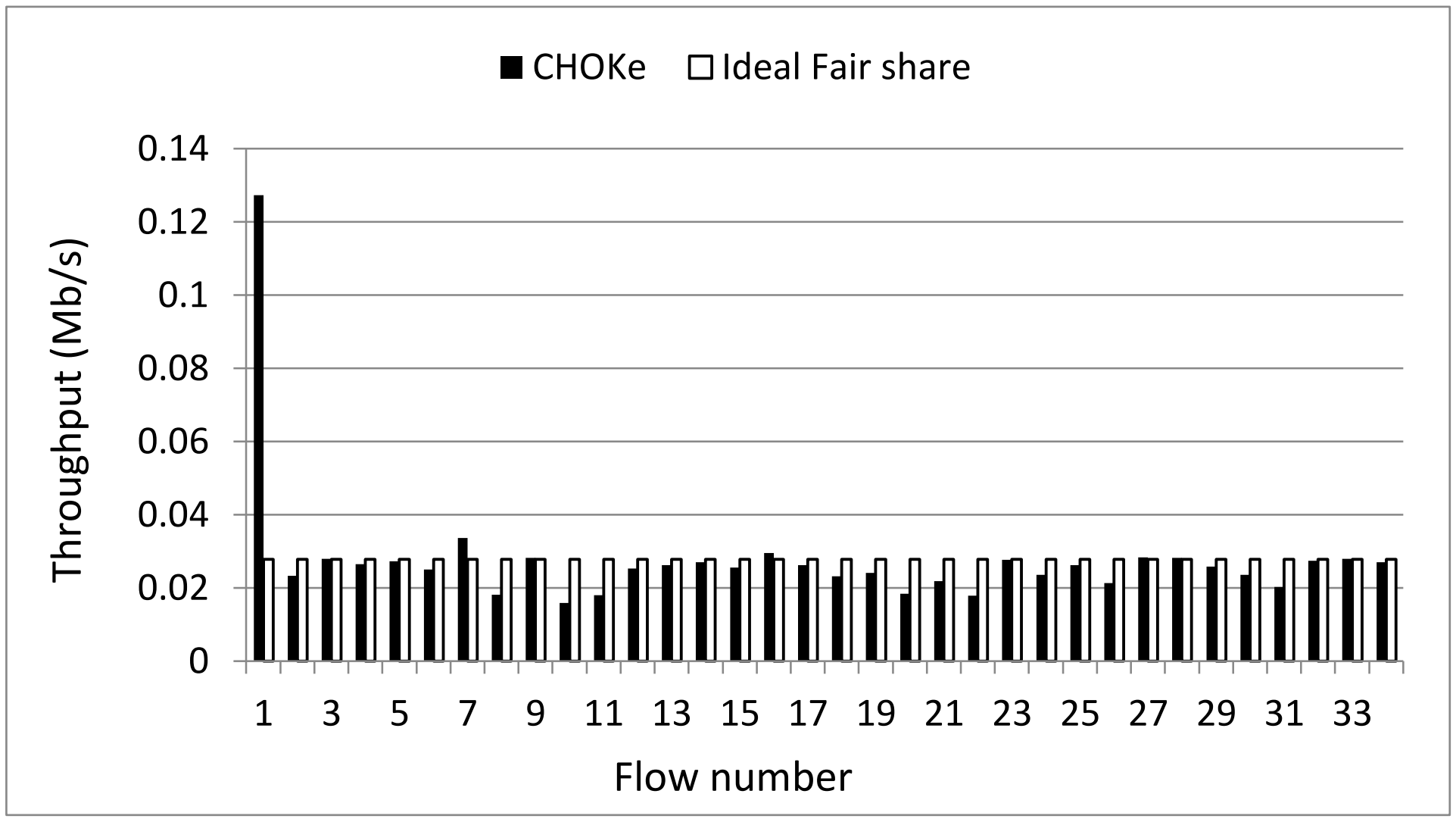

FIG. 3.3 THROUGHPUT UNDER CHOKE

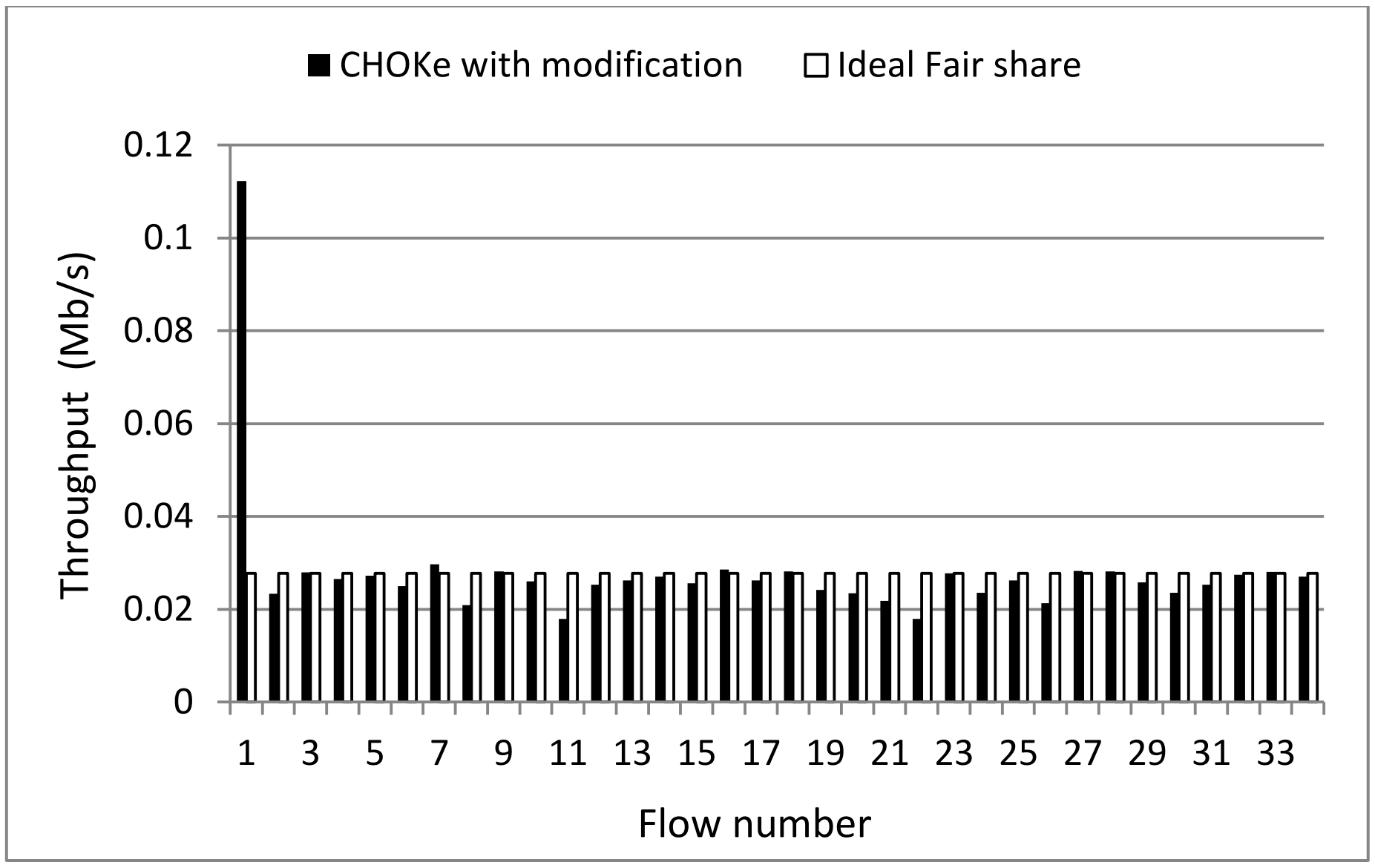

FIG. 3.4 THROUGHPUT UNDER CHOKE WITH MODIFICATION

2. Although CHOKe claims fairness, but its fairness significantly declines under many unresponsive flows, because of its static drawing factor rather than increasing drawing factor, *i.e.*, more punishment is needed as queue getting full with unresponsive flows.

### 3.1.2 GCHOKE'S SHORTCOMINGS

gCHOKe performs better than the CHOKe but gCHOKe does not addresses the problems of CHOKe and generalizes its drawing factor. Therefore, gCHOKe has all the drawbacks of CHOKe.

gCHOKe has following drawbacks:

1- gCHOKe's drawing candidate packet criteria is same as of original CHOKe so it also lacks information about arriving packet and drawing packet. Since more packets of malicious flows accumulated at the rear end, gCHOKe doesn't show concerns about this.

2- gCHOKe fails to outperform CHOKe in the presence of many malicious streams as "*maxcomp*" fails to trigger more comparisons due to wrong packet selection criteria. Thus, in the presence of many unresponsive streams, gCHOKe behaves like original CHOKe.

#### 3.1.3 SOLUTION FOR CHOKE AND GCHOKE'S SHORTCOMINGS

The proposed AQM scheme CHOKeD (CHOKe Descendant) is completely stateless approach to ensure fairness among all flows of a router. To minimize the above described shortcomings of CHOKe and gCHOKe, CHOKeD proposes location based dynamic drawing factor. This thesis suggests that the numbers of candidate packets are dynamic, on the arrival of a packet at the queue. In the market, there are different sized buffers are available consequently AQM algorithm should be scalable for cooperating these sizes. If the buffer size and current queue size changes, CHOKeD updates its drawing factor accordingly.

## 3.2 DESIGN GOALS OF CHOKED

Factors like Complexity, Queuing Delay, Congestion Indicator, Congestion Control Function and Feedback to End Hosts are studied here before designing of a new AQM protocol following requirements and basic design goals are considered in the thesis.

### 3.2.1 COMPLEXITY

Computation complexity is regarded as a vital element for performance assessment of the AQM protocol in terms of time and memory. According to network design principle, network backbone must be simple and if complexity has to be added then must be added at end hosts only (Abbas, *et al.*, 2001). The complexity at the network router means more delay for a single packet processing, so complex AQM based solutions are discouraged because of their complexity for real-time traffic. Complex systems are practically not a good solution for physical deployment over network core, so network core much be kept simple and flexible (Ku *et al.*, 2005). Thus, only simple and effective AQM schemes are considered as a better choice for practical implementation. The design goal of CHOKeD is to develop an AQM algorithm with minimal complexity.

### 3.2.2 Queuing Delay

It is routine to deploy AQM schemes over routers of various sizes practically. Queue delay is the time that a packet spent in queue before its transmission. In router based AQM, packet transmission time increases as the size of router increases. Large routers cause more delay for a single packet transmission, due to this reason it is considered as overhead for network traffic. As large queue buffers are reason for more queuing delays, that is why large queue buffers are not suitable for practical implementation (Adams, 2013). The design goal of CHOKeD is to reduce queuing delays.

### 3.2.3 Congestion Indicator

AQM scheme become operational, over network congestion indication. On receiving a congestion indication, AQM protocol must perform certain actions to overcome this congestion. Usually AQM schemes detect incipient congestion and become active for incoming traffic. Thus, an AQM scheme must have an indicator that detects incipient congestion (Adams, 2013). RED, CHOKe and gCHOKe uses average queue length as congestion indicator. The design goal of the proposed AQM scheme is to have a congestion indicator to support its congestion control mechanism.

### 3.2.4 CONGESTION CONTROL FUNCTION

After successful congestion detection, AQM protocol must have to initiate its congestion control function or mechanism. Based on the level of congestion, AQM protocols perform actions of queuing or deleting packet to control congestion (Adams, 2013). AQM scheme must be active on receiving congestion indication and perform operations of queuing or deleting the newly arrived packets.

### 3.2.5 FEEDBACK TO END HOSTS

End hosts reduce or increase their transmission rates, on arrival of feedback from router-based AQM mechanism. That is why TCP/AQM is distributed system. To notify end hosts about the level of congestion, packets can be ECN (Explicit Congestion Notification) marked or dropped from the queue. Moreover, AQM systems that use dropping are more stable than ECN enabled (Deng *et al.*, 2002). To notify end hosts, the design goal of CHOKeD is to drop some packets due to ongoing congestion. The dropper packets will serve as a feedback mechanism in CHOKeD. However, ECN-enabled feedback mechanism can be incorporated in CHOKeD.

### 3.2.6 FAIRNESS AND COMPLEXITY

Fairness among all flows of the queue is a basic objective of any AQM system. Fairness can be implemented through various solutions or methods. Optimized approaches are less complex in terms of time and state complexity, but higher in fairness. Ultimately AQM protocol ensures fairness among all flow, but there is a tradeoff between simplicity and superior fairness (Kamra *et al.,* 2000). Thus, with simpler but practically deployable approach only up to some level of fairness can be achieved. The design goal of CHOKeD is provide high fairness with less complexity.

#### 3.2.7 RANDOM PACKET DROPPING

AQM systems use different approaches for fairness implementation over network routers. These methods are either sure packet dropping or random packet dropping. Sure, dropping are not encouraged solution as only there is going to be congestion (incipient congestion) rather than link become congested. Mostly, AQM approaches adopt random dropping because random packet dropping improves fairness (Li *et al.,* 2001). CHOKeD adopts random packet dropping to notify end hosts.

Later sections describes the complete mechanism of the proposed novel AQM scheme, CHOKeD.

## 3.3 CHOKeD (CHOKe DESCENDANT)

There is a need to develop a robust fairer AQM solution that provides better protection for TCP flows as compared with CHOKe and gCHOKe. The proposed scheme CHOKeD is completely stateless AQM approach as it needs no record keeping for number of flows residing in the queue. CHOKeD is motivated by CHOKe (Pan *et al.,* 2000) which is a much better approach than traditional AQM approaches RED and Drop Tail (Floyd and Jacobson, 1993) to prevent friendly flows like TCP from unfriendly or unresponsive flows like UDP flows.

Consider a router that maintains a single FIFO buffer for the packets of all the flows that sharing an outgoing link. CHOKeD methodology is based on three components:

1. Queue regions
2. Drop candidate packet's selection from queue, *i.e.,* called dynamic drawing factor, $D_i$
3. Location of drop candidate packets from the queue.

CHOKeD's congestion indication is same as of RED, *i.e.*, average queue size $Q_a$ Eq. 3.1 shows the expression of average queue size.

$$\mathrm{Q_a} \leftarrow (1 - \mathrm{w_q})\mathrm{Q_a} + \mathrm{w_q}\mathrm{Q_c} \tag{3.1}$$

Where, $w_q$ is queue weight, same as in RED, while $Q_c$ is current queue size. The reason behind the selection of average queue size as congestion indications is that its value does not change much as compared with $Q_c$. CHOKeD wants to be a stable system under current queue size oscillations that is why it has average queue size as congestion indication.

Unresponsive flow with high-bandwidth is likely to have more packets in the buffer, hence a higher probability for flow matching and consequently dropping. Due to this, CHOKeD increases its drop candidate packets for match-drop comparisons to punish the unresponsive flows and restricting them from capturing the whole buffer space.

When a packet $pk$ arrives at congested buffer, the average queue size $\mathrm{Q_a}$ is calculated as in Eq. 3.1, and CHOKeD performs the following actions that are shown in Fig 3.5.

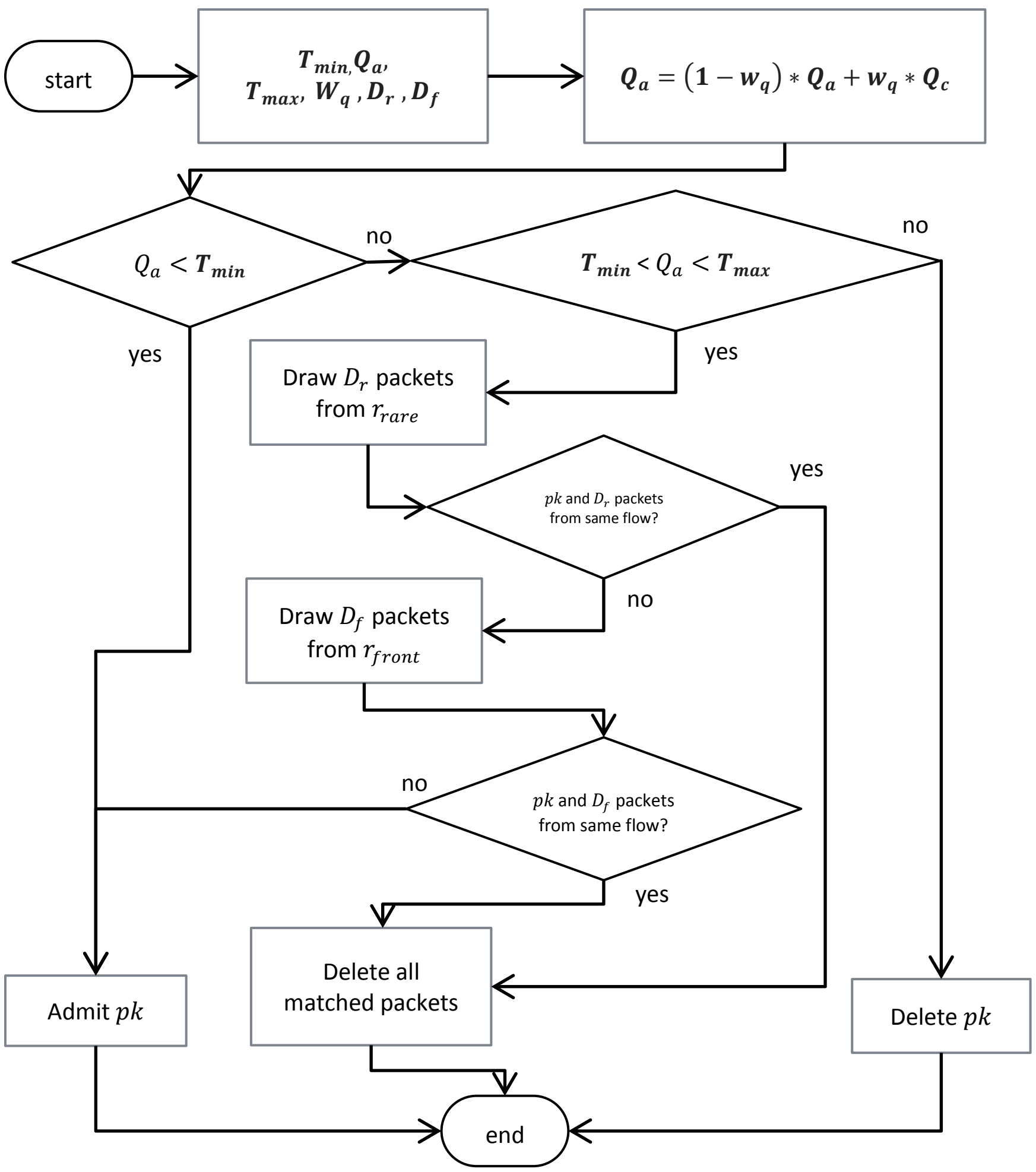

FIG.3.5 CHOKED FLOWCHART

First of all, the input parameters, average queue size thresholds, *i.e.*, minimum threshold $T_{min}$, maximum threshold $T_{max}$, queue weight $w_q$ and average queue size $Q_a$, are initialized. If $Q_a$ is less than the $T_{min}$, CHOKeD admits newly arrived $pk$. If $Q_a$ is greater than the $T_{max}$, CHOKeD does not allow packet $pk$ to enter the queue. Otherwise, if the average queue size is greater than the $T_{min}$ but less than the $T_{max}$, CHOKeD slices the queue into two equal size regions, named as rear and the front regions. Firstly, it draws $D_r$ packets from the rear queue region and match with the flow id of $pk$. The value of $D_r$ is determined by Eq. 3.3. If all or any of the packets from the queue has same flow id, same as of the $pk$, CHOKeD deletes all matched packets and also $pk$. In this comparison, if none of candidate packets from the queue has the same id as that of $pk$, CHOKeD returns all $D_r$ packets back to the queue and draws the $D_f$ packets from the front region for comparison, before admitting $pk$. The value of $D_f$ is computed using Eq. 3.4. Now CHOKeD compares, $D_f$ candidate packets, drawn from the front queue region, with flow id of $pk$, if any or all packets have same flow id as of $pk$, packet $pk$ and all matched packets are dropped. In the event of no matching at this moment, the candidate packets are restored back to the queue, but the $pk$ may still be dropped with a probability that depends on the level of congestion. Detailed algorithm is discussed in appendix A.

### 3.3.2 Queue Regions

Initially, on arrival of a packet, CHOKeD divides the current queue occupancy $Q_c$ into two dynamic regions of equal sizes, rear $r_{rear}$ and $r_{front}$ front respectively shown in Fig 3.6.

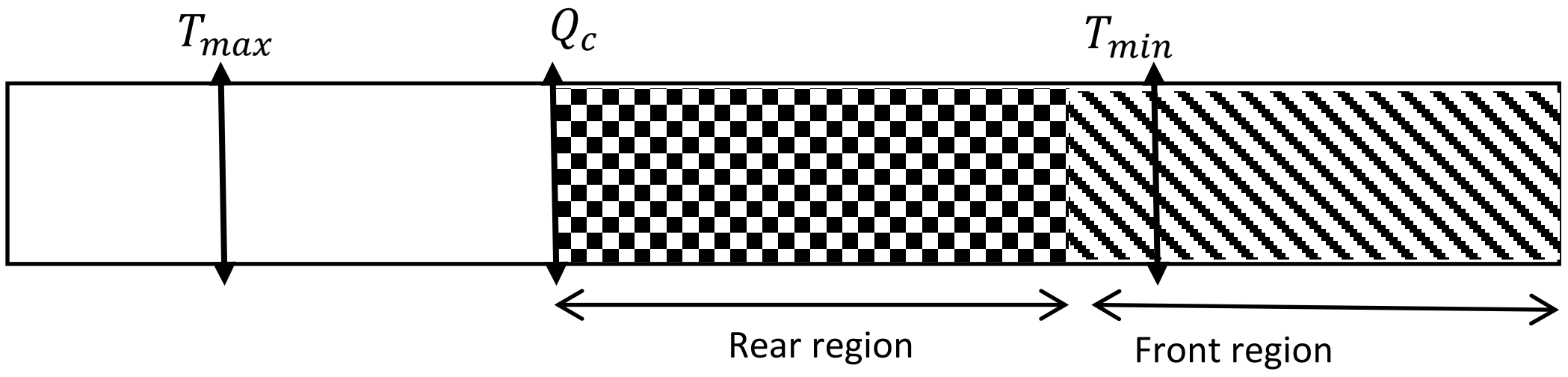

Fig. 3.6 Queue regions, on arrival of a packet

If $Q_c$ is the current queue size, then on arrival of a packet from any flow $i$ $(i = 1,2,3,\ldots,n)$, CHOKeD slices the queue into the rear and front regions. Reason behind explicit two regions is that CHOKeD want to filter or punish misbehaving flows that are just arrived at the queue. Also region portioning helps to protect responsive flows from unresponsive flows. As in section 3.1.1, experimentally this confirms that the numbers of unresponsive flows are more accumulated at rear queue region, so there is need to handle the rear region and the front queue region distinctively.

### 3.3.3 DRAWING FACTOR

Basically drawing factor is responsible for controlling of unresponsive flows and maintaining fairness. CHOKeD is based on the idea that packets should be drawn in case of increasing current queue size, $Q_c$ and fewer in the case of decrease in $Q_c$. After slicing the current queue logically, $D_i$ (where, $i = f, r$) is calculated and $D_i$ packets are drawn from the queue, on the arrival of every new packet. Drawing of $D_i$ packets form queue depends upon two parameters: (1) current queue size (2) Size of router's buffer $B$. Candidate packets from the queue are determined through the Eq. 3.2, while for the rear region, $D_r$ and for the front queue region $D_f$ expressions are given in Eq. 3.3 and Eq. 3.4, respectively.

$$D_i = Round\left(\frac{(Q_c * \sqrt{B})}{(T_{max} - T_{min}) * ln(B)}\right) \tag{3.2}$$

$$D_r = D_i \tag{3.3}$$

$$D_f = Round\left(\frac{D_i}{2}\right) \tag{3.4}$$

CHOKeD does not introduce any extra variables and just utilizes already available parameters from the system such as current queue size and buffer size. Eq. 3.2 illustrates that the value of $D_i$ increases if value of current queue size $Q_c$ and buffer size $B$ increases.

There are two main reasons for selection of such drawing factor for CHOKeD:

1. Firstly, increase in the queue size increases the chances of unresponsive flow packets in the queue. Consequently, there is a need to increase in the value drawing factor $D_i$ accordingly.
2. Secondly, router's buffer size $B$ varies from vender to vender, large sized buffers will allocate more space for traffic, resulting more unresponsive flow packets in the queue and therefore, there is also need to update the drawing factor according to the buffer size. Thus, CHOKeD is conformable to all buffer sizes.

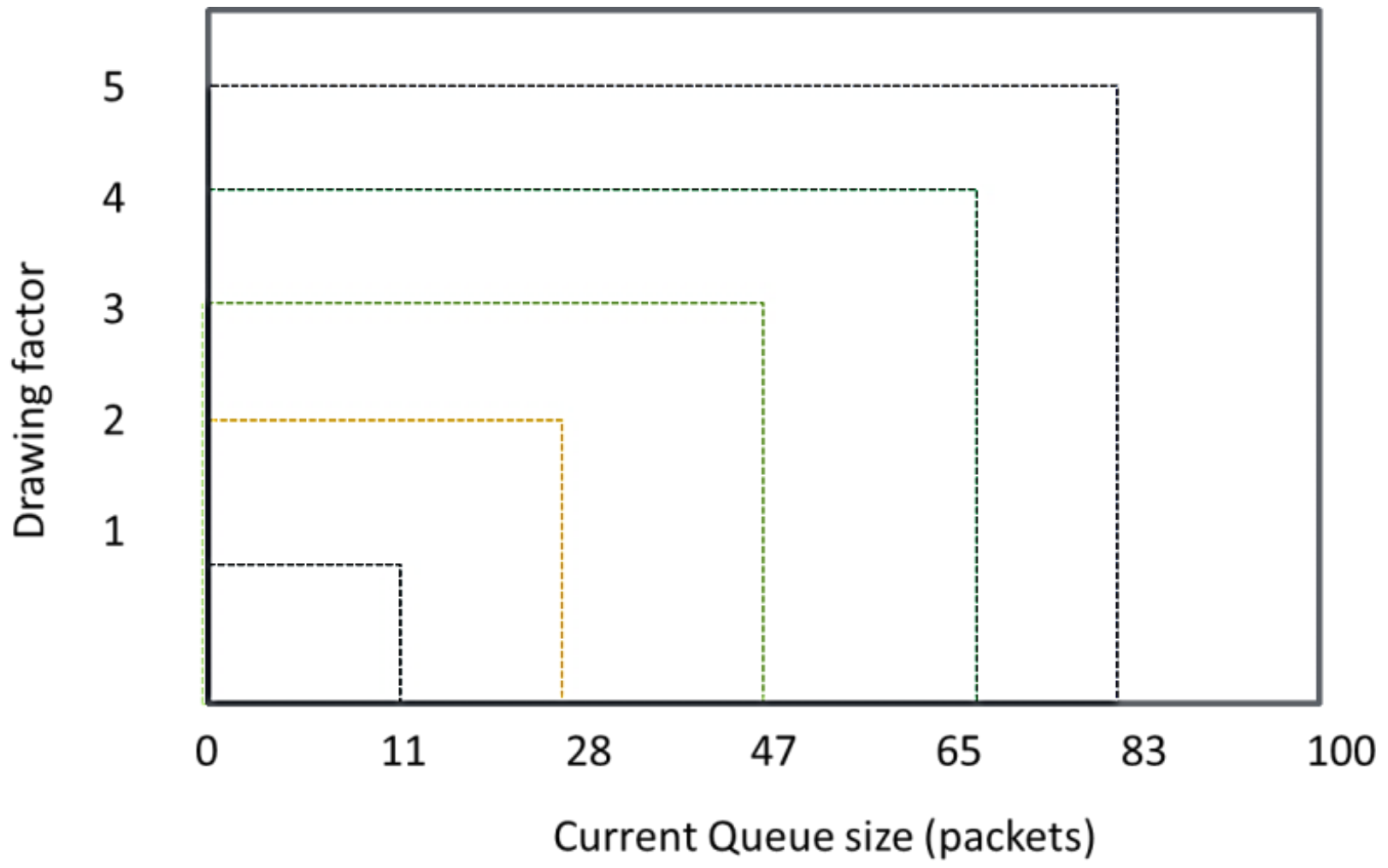

FIG.3.7 CHOKED DRAWING FACTOR

Fig. 3.7 shows the behavior of Eq. 3.2, Fig. 3.7 illustrates an increasing trend of $D_i$, due to increase in the current queue size. Here it is assumed $B$ contains 100 packets. For a buffer of size 100 packets, $Q_c$ can have maximum 100 packets in it, the maximum value of $D_i$ is 5 on 83 packets in $Q_c$ according to Eq. 3.2. In case, there are 11 packets in the $Q_c$ only, value of $D_i$ is 1. CHOKeD drawing factor increases linearly as current queue size increases. If queue holds 83 packets, CHOKeD draws 5 packets from the rare region. If this comparison is unsuccessful, then CHOKeD draws 3 packets from the front queue region.

CHOKeD chooses a dynamic number of drawing candidate packets from the queue, because if $Q_c$ continue to increase again and again towards full queue, eventually, the average queue size increases, this indicates that the queue is going to be congested soon. As misbehaving flows are ignorant of ongoing congestion and continue transmission unresponsively, so they occupy more buffer space than from responsive flows. Thus, punishment factor increases linearly to punish more vigorously misbehaving flows.

# CHAPTER 4

## 4 SIMULATIONS AND RESULTS

This chapter rigorously evaluates the proposed CHOKeD algorithm. Several performance metrics are used to evaluate the efficiency of the proposed AQM protocol, in regulating malicious or unresponsive flows. These metrics include throughput, queuing delay, goodput, fairness and other metrics like TCP with different RTT, TCP inter-protocol fairness and web-traffic. Moreover, different traffic scenarios are used in this thesis, to depict the real internet behavior and shown in Table I. All simulations are carried out with the help of Network Simulator (NS) 2, in short NS-2, with version 2.35. For simulations, dumbbell topology is used and this type of topology is also used by authors of CHOKe (Pan *et al.*, 2000) and gCHOKe (Eshete and Jiang, 2013). Fig. 4.1 shows dumbbell topology with TCP and UDP sources.

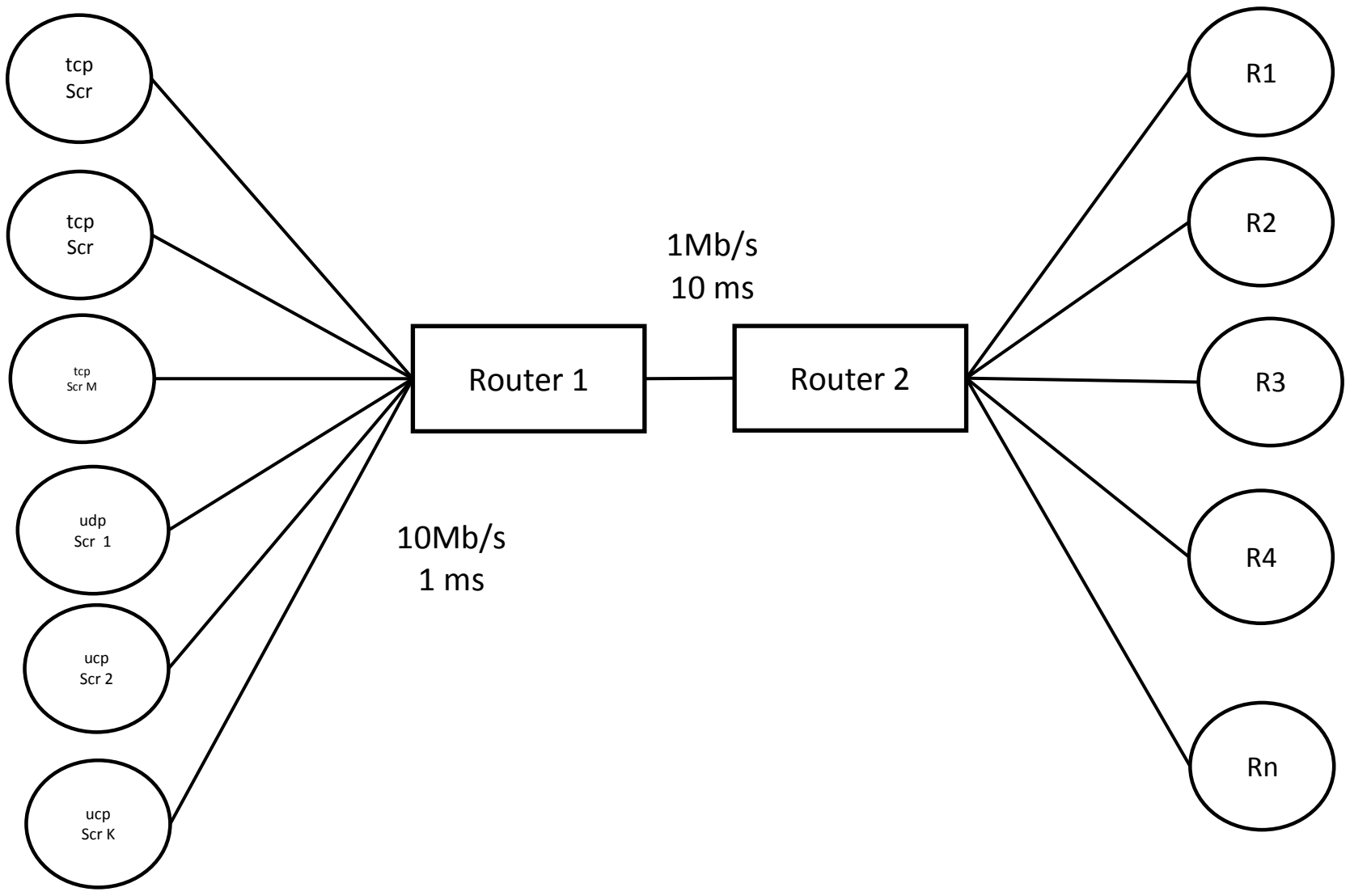

FIG 4.1: BOTTLENECK DUMBBELL TOPOLOGY

TABLE I. EXPERIMENTAL MODEL FOR SINGLE AND MULTIPLE TRAFFIC STREAMS

| **Model** | **Number of nodes** | **TCP flow** | **UDP flow** | **Outgoing link capacity** |
|---|---|---|---|---|
| 1 | 34 | 33 | 1 | 1 Mb/s |
| 2 | 25-100 | 22-88 | 3-12 | 1 Mb/s |

From Fig. 4.1, sender nodes at left side, receiver nodes at right while a bottleneck link in the center. Sender nodes have TCP and UDP flows. In this topology, buffer size $B$ is 100 packets, link capacity $C$ is 1 Mb/s with 10 ms delay and average packet size is 1 Kbytes. All TCP sources can transmit at the rate of 10 Mb/s. TCP flows are driven by FTP applications and UDP flows are driven by CBR traffic at the speed 2 Mb/s, twice than the link capacity. To compare CHOKeD with RED, CHOKe and gCHOKe these AQM parameters are used, as described by (Pan *et al.*, 2000), average queue size minimum threshold $T_{min} = 40$ packets, average queue size maximum threshold $T_{max} = 80$ packets, and queue weight $w_q = 0.02$.

Two types of traffic models are used to validate the proposed CHOKeD (1) multiple responsive flows and single unresponsive flow (2) multiple responsive and multiple unresponsive flows and shown in Table I.

## 4.1 NETWORK SIMULATOR 2 (NS-2)

NS-2 is a discrete event simulator, developed for Internet networks research. For both types of networks, wired and wireless, NS provides extensive support for simulation of TCP, routing, queue management, traffic modeling and multicast protocols. Actually NS began as a variant of the "REAL network simulator" in 1989 and in the past few years it is evolved significantly. Basic requirements of NS-2 are a computer and a C++ language compiler. NS-2 could be installed on Linux, SunOS, and Solaris operating systems. Through Cygwin package NS-2 can be installed on Windows operating system. NS-2 allows researchers to install it via all-in-one package or through component by component installation. NS-2 uses Tcl (Tool Command Language) for scripting and C++ for protocol designing. Xgraph is used to draw graphs for evaluations of the underlying protocol. "awk" language is used to read trace files of NS2.

## 4.2 MULTIPLE RESPONSIVE AND SINGLE UNRESPONSIVE FLOW

CHOKeD is endorsed against RED, CHOKe and gCHOKe as illustrated in first simulation is considered traffic model 1, *i.e.*, 33 TCP responsive flows and one unresponsive UDP flow. Through simulations, flow throughput, goodput, queuing delay and fairness are computed. All these performance metrics are discussed in later sections.

### 4.2.1 THROUGHPUT ANALYSIS

The number of bits that can be transferred over a period of time is network's throughput (Peterson *et al.*, 2007). By means of throughput, the system can be evaluated with best link utilization and fairness among all the flows. High throughputs of responsive flows indicate good performance the AQM protocol. Similarly, lower the throughput of unresponsive flow, better the performance of AQM protocol (Adams, 2013). This simulation measures, TCP and UDP throughputs separately for all comparison techniques including CHOKeD.

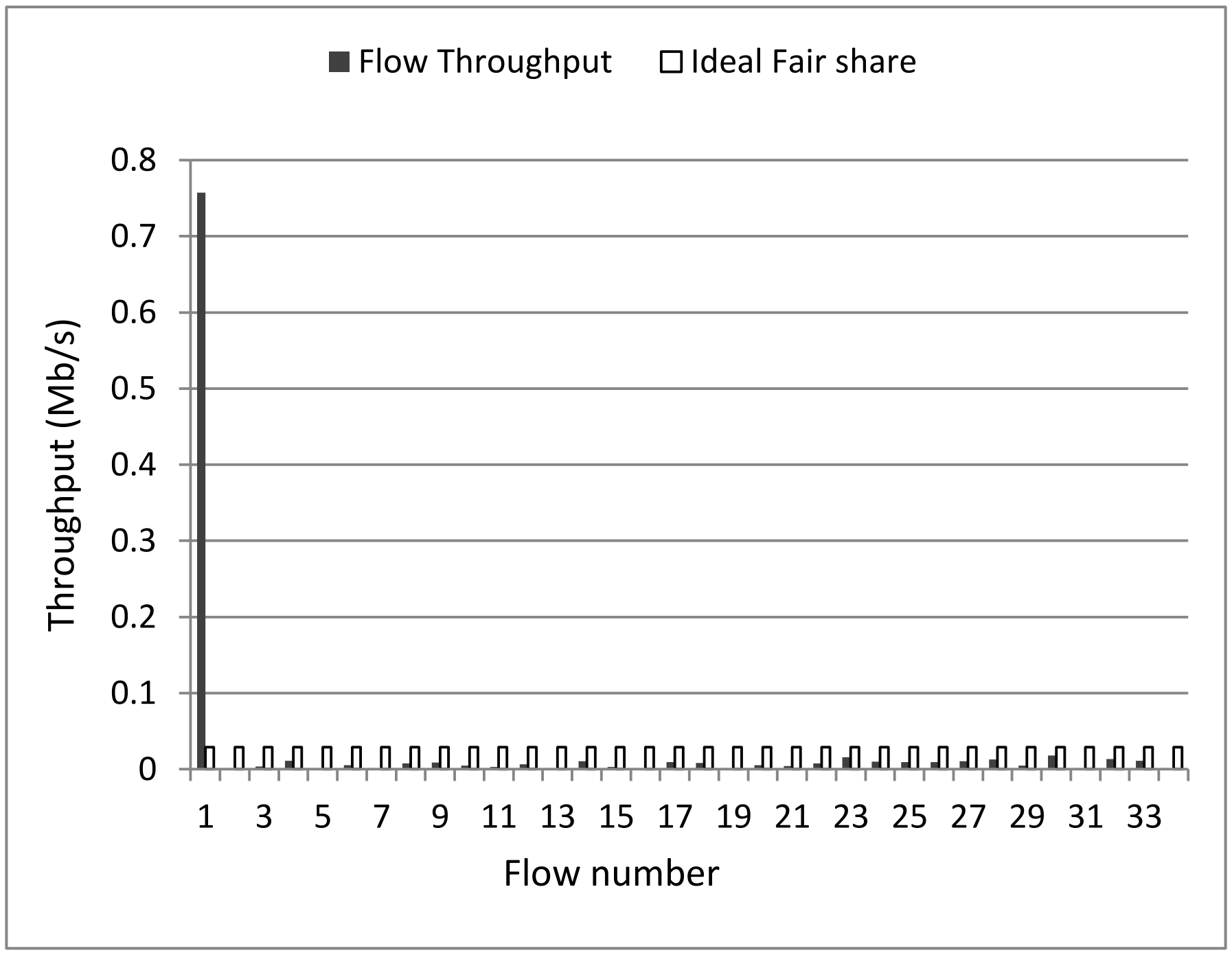

FIG. 4.2 TCP AND UDP THROUGHPUT, UNDER RED

The trends in Fig. 4.2 confirm the throughput graph of RED with 33 TCPs and 1 UDP flow. Flow 1 is UDP, while Flow 2 to Flow 34 are TCP flows. Simulation graph also has ideal fair share of flows. Under RED, throughput of UDP flow has highest value with its ideal fair share. This highest value of UDP throughput, *i.e.*, 0.75 Mb/s, indicates that unresponsive UDP steals the major portion of the link bandwidth and cause starvation of TCP flows, as they only get a minor portion of the link bandwidth. This is a complete failure of RED under 34 flows with a single UDP unresponsive. On the other hand all TCP flows even near to cutoff. Thus, RED fails in the presence of unresponsive flows.

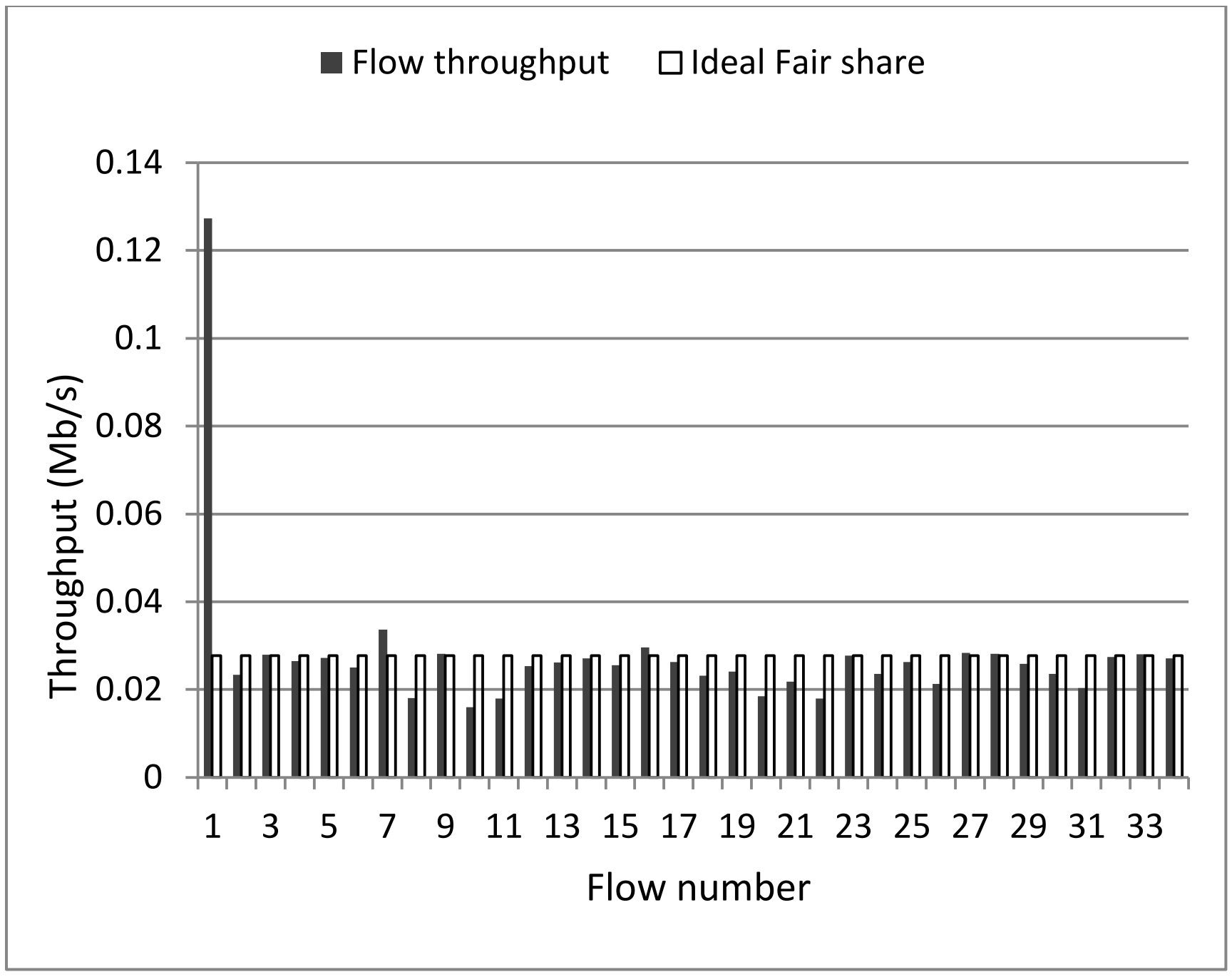

FIG. 4.3 TCP AND UDP THROUGHPUT, UNDER CHOKE

The results in Fig 4.3 reveal throughput of 34 flows verses their flow number, under the queue managed by CHOKe. In this case, most of TCP flows get more bandwidth fair share as compared with RED's TCPs. But still single UDP flow's throughput, *i.e.*, 0.12 Mb/s is still higher than all TCP flows throughput and also higher from its fair share that is 0.029 Mb/s. Under CHOKe unresponsive flow still get high throughput that is why CHOKe fails to behave like a fair bandwidth allocation algorithm.

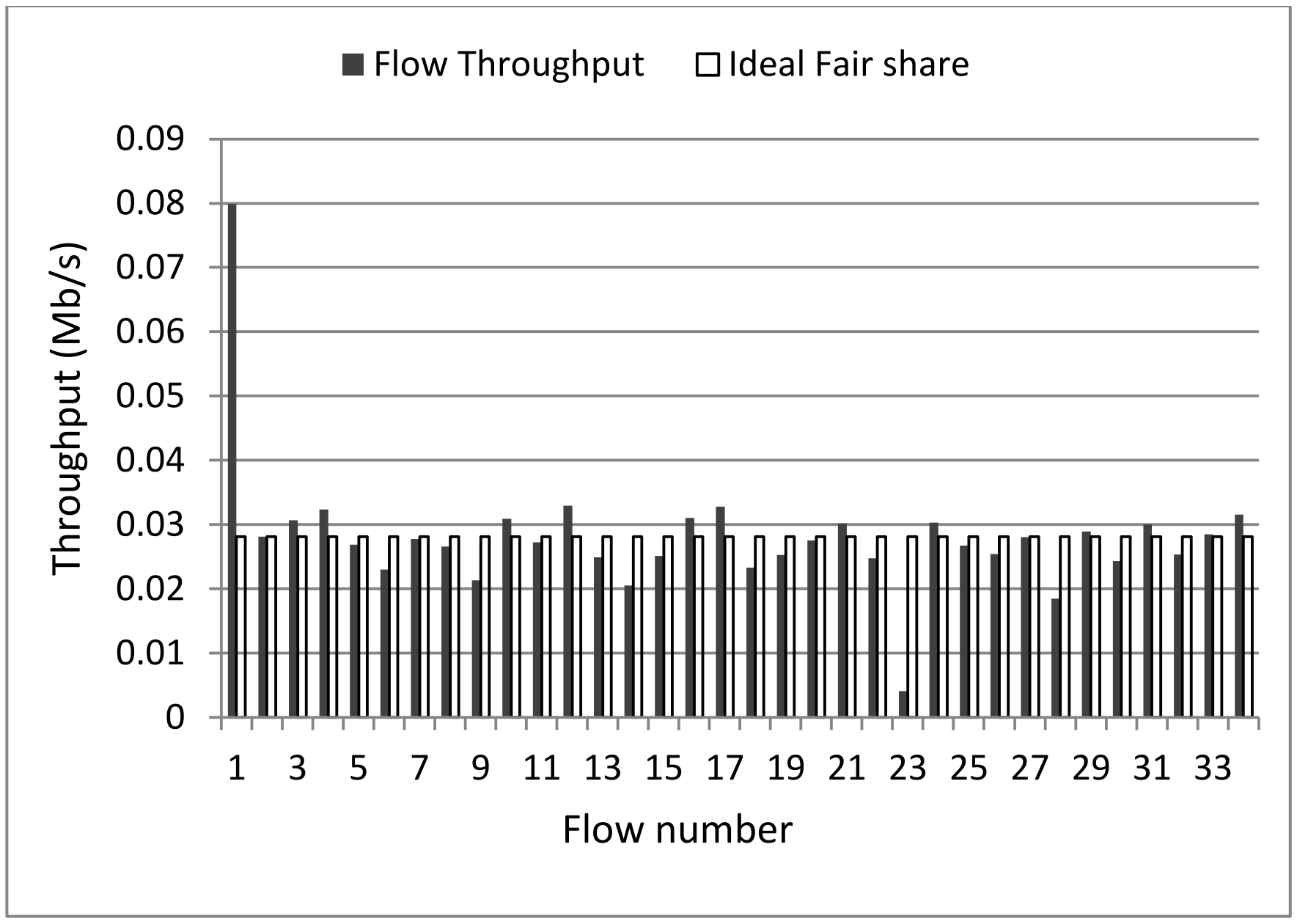

FIG. 4.4: UNDER GCHOKE THROUGHPUT OF 33 TCP AND SINGLE UDP

The results in Fig. 4.4 show TCP friendly behavior of gCHOKe. gCHOKe still fails to restrict UDP flow up to its fair bandwidth share. UDP successfully steals high buffer share under gCHOKe too. Somehow gCHOKe performance is still better than RED and CHOKe but not fair yet. Only few TCPs get reasonable fair shares. The higher peak value of UDP throughput, *i.e.*, 0.08 Mb/s as far greater than 0.029 Mb/s its fair share indicates gCHOKe's less control over unresponsive flow. gCHOKe's bandwidth allocation of all flows is better than RED and CHOKe.

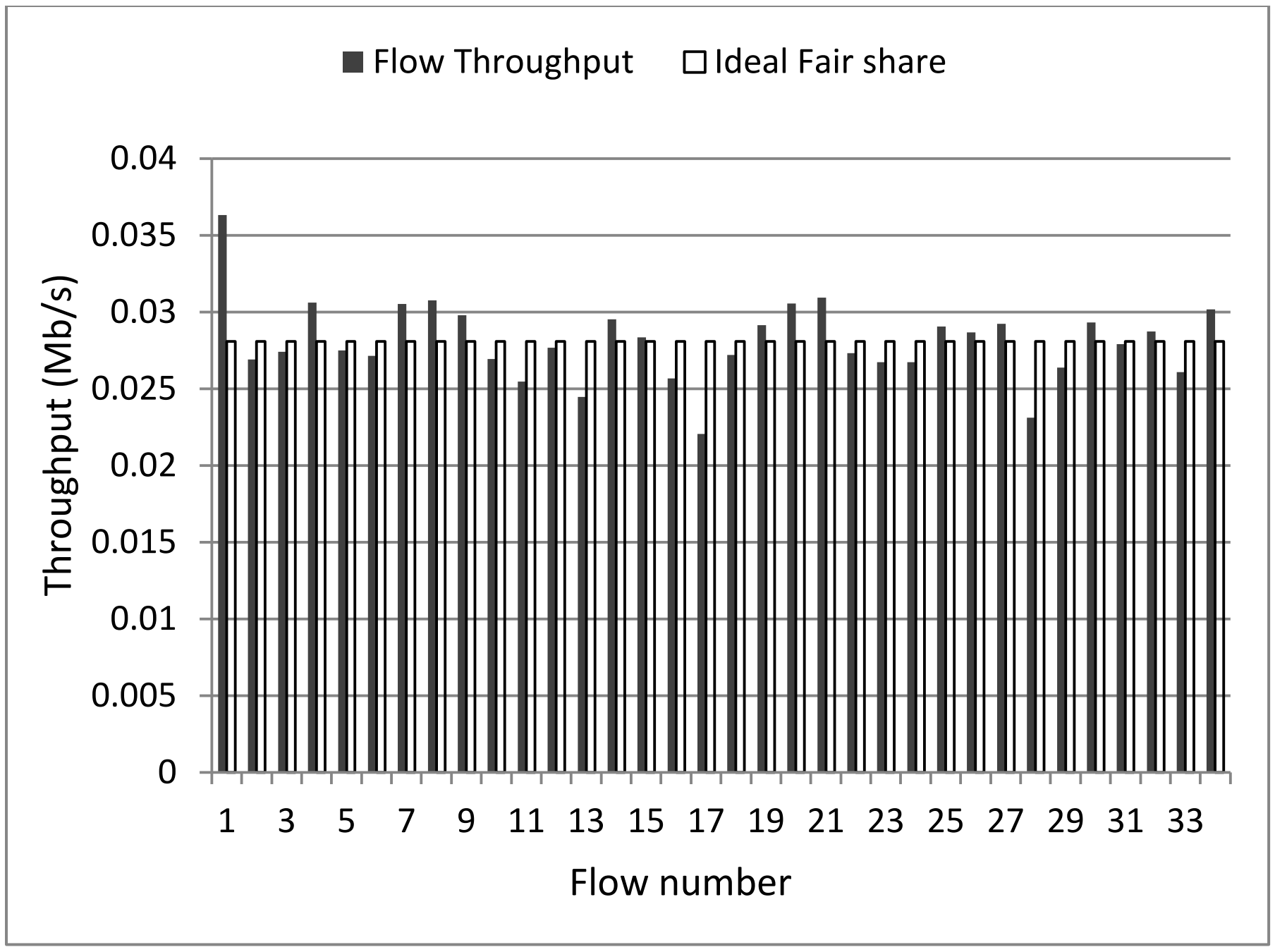

FIG. 4.5: UNDER CHOKeD THROUGHPUT OF 33 TCP AND SINGLE UDP

The results in Fig 4.5 reveal throughput of 34 flows verses their flow number, under the queue managed by CHOKeD. Firstly, none of TCP flow is near to cutoff and most of TCP flows easily achieve their fair bandwidth share. Moreover, in this simulation, UDP throughput, *i.e.*, 0.036 Mb/s, is nearly equal to its bandwidth share, *i.e.*, 0.029 Mb/s, in contrast with UDP throughput under the queue managed by CHOKe or gCHOKe. Form the graph it is clear that CHOKeD successfully penalizes unresponsive flow and protects TCPs. Thus, CHOKeD outperforms RED, CHOKe and gCHOKe.

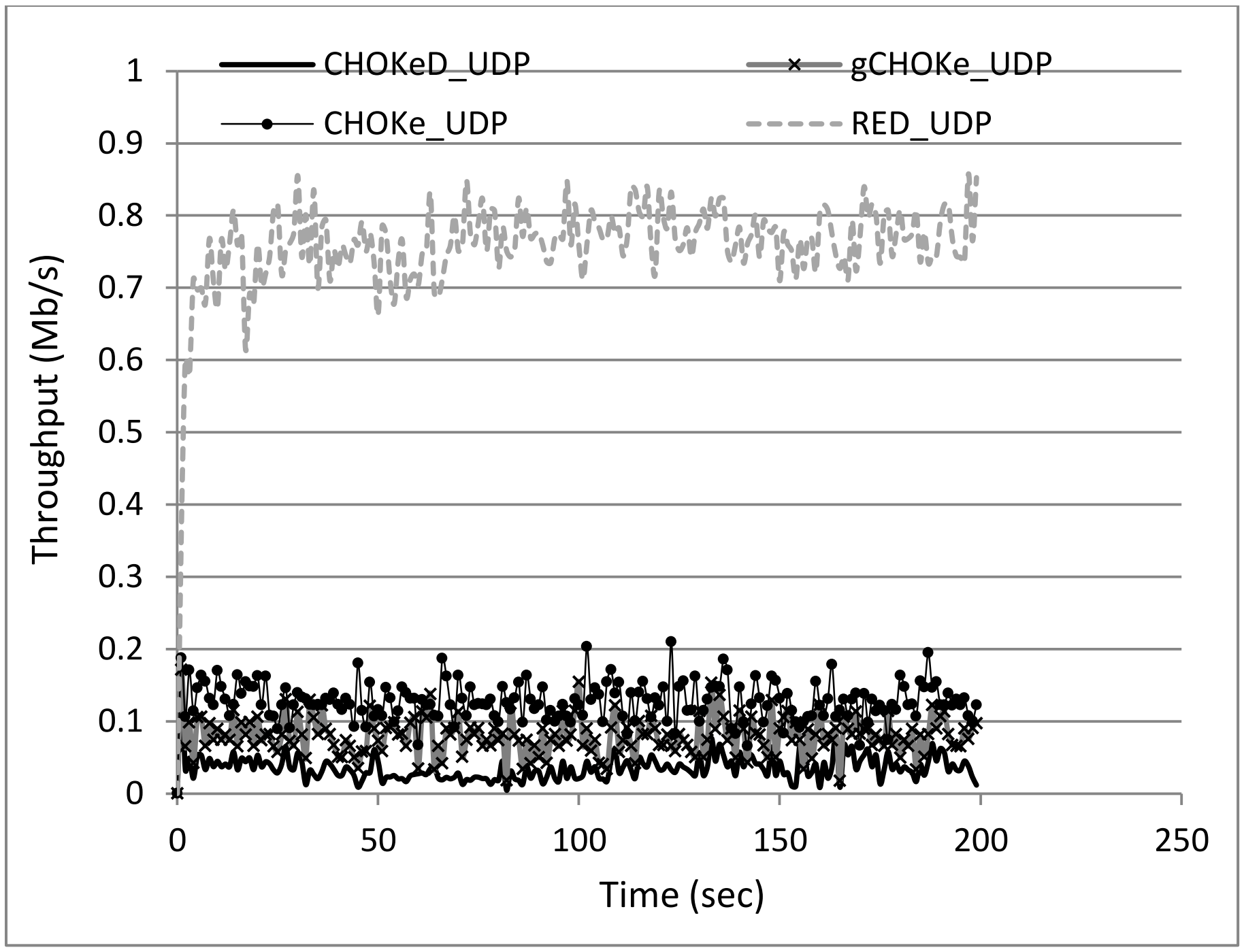

FIG. 4.6-A: UDP THROUGHPUT OF RED, CHOKE, GCHOKE AND CHOKED

The results in Fig. 4.6-A and 4.6-B show throughputs of UDP flows verses execution time under 33 TCP and 1 UDP flows. Fig 4.6-A shows the throughput of UDP flow under RED, CHOKe, gCHOKe and CHOKeD while Fig 4.6-b shows the throughput of UDP flow under CHOKe, gCHOKe and CHOKeD. Under RED UDP has very higher value than its ideal fair share. CHOKe and gCHOKe cannot restrict UDP near the ideal fair share. While only CHOKeD manages to allow UDP up to its fair share level. Consequently, only CHOKeD maintains unresponsive flow under tight control.

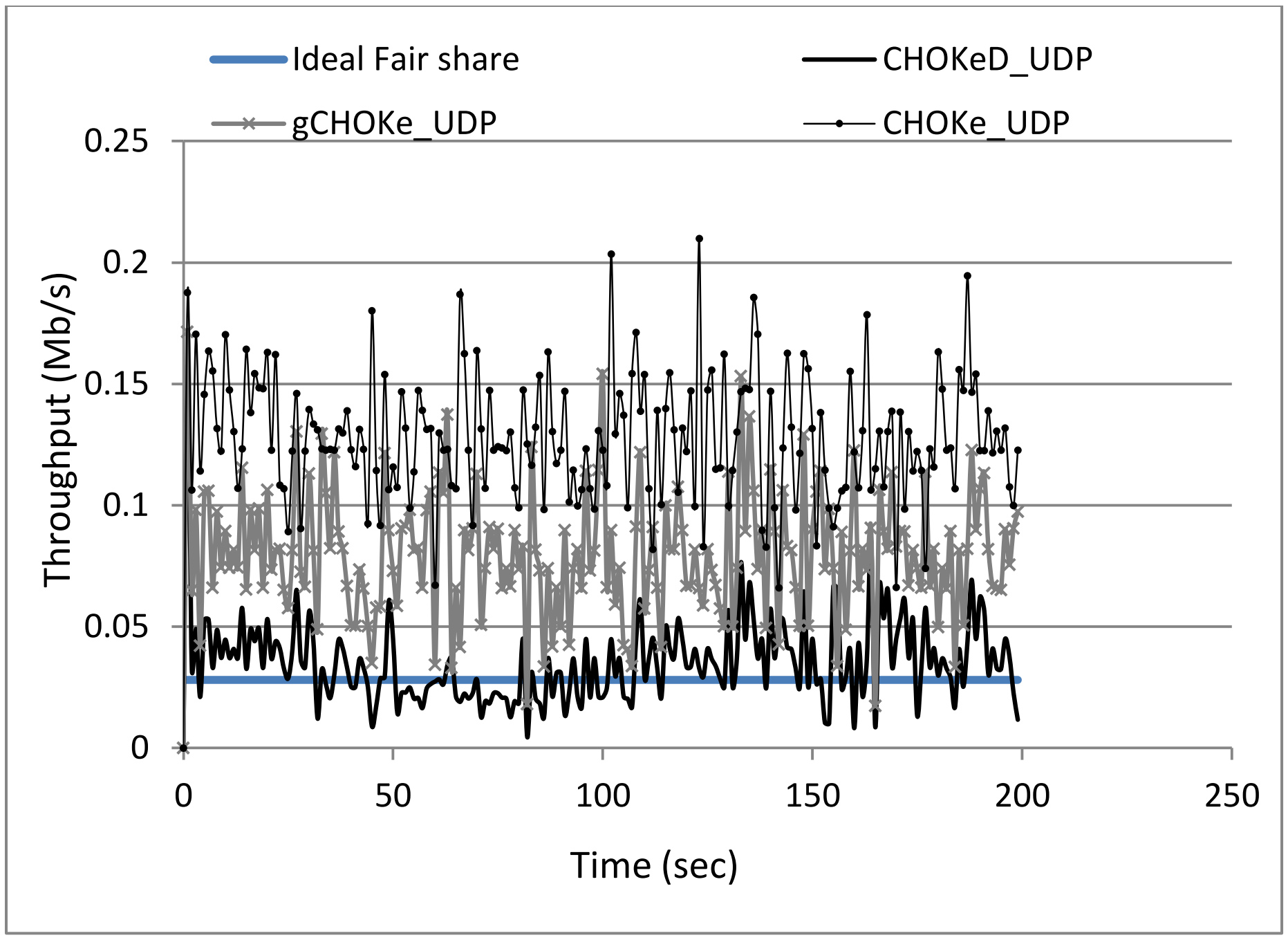

FIG. 4.6-B: UDP THROUGHPUT OF CHOKE, GCHOKE AND CHOKED

### 4.2.2 GOODPUT, FAIRNESS AND QUEUING DELAY ANALYSIS

Low *et al.*, (2004) defined goodput as *"The total amount of effective data that transmitted through the network"*. Actually, here effective data means non-replicated data. Mathematically TCP and UDP goodput are expressed in Eq. 4.1 and Eq. 4.2.

$$Goodput(TCP) = \frac{R_x - T_x}{T} \tag{4.1}$$

$$Goodput(UDP) = \frac{R_x}{T} \tag{4.2}$$

Where, $R_x$ and $T_x$are received and retransmitted bits respectively. $T$ denotes the duration of flow.

In 1984 Jain, proposed an index value as a fairness indicator for all flows under congested link. This this uses, Jain's fairness index to determine CHOKeD's fairness under bottleneck. Jain's index calculated through Eq. 4.3, (Jian, 1984). Where, $0 < \mathrm{f(x)} \leq 1$ .

$$f(x) = \frac{(\sum_{i=1}^{N} x_i)^2}{(n\sum_{i=1}^{N} x_i^2)} \tag{4.3}$$

The closer the value of fairness index is to 1, better the fairness. Jain fairness index values are shown in Table II.

Queuing delay is used to evaluate the performance of router-based AQM schemes. *"*Queue delay is the time that a packet spent in the queue before it is transmitted*"*. Schemes with lower queuing delay are encouraged to implement on routers (Adams, 2013). Eq. 4.4, shows mathematical expression of queuing delays, where $T_a$and $T_s$ are arrival and sending time from the buffer, while $N_f$ represents number of flows at a given time.

$$D_Q = \sum(T_a - T_s)/\sum N_f \tag{4.4}$$

Using Model 1 of Table I, goodput, fairness and queuing delay results are shown in Table II. Form Table II, it is learnt  that TCP successfully attains high goodput, *i.e.,* 0.87 Mb/s as compared with CHOKe having 0.77 Mb/s, gCHOKe 0.82 Mb/s and RED having 0.21

Mb/s. This high goodput shows TCP-friendly behavior of CHOKeD. Simulation results verify the effective behavior of CHOKeD protecting TCP with higher goodput.

TABLE II. GOODPUT , FAIRNESS AND QUEUING UNDER AQM SCHEMES

| **AQM** | **TCP Goodput (Mb/s)** | **Fairness** | **Queuing Delay (*s*)** |
|---|---|---|---|
| RED | 0.211162 | 0.5357 | 0.363512 |
| CHOKe | 0.77096 | 0.70999 | 0.298696 |
| gCHOKe | 0.826655 | 0.87919 | 0.275913 |
| CHOKeD | 0.878491 | 0.9668 | 0.233061 |

In the presence of single UDP flow, RED shows 53% fairness while CHOKe and gCHOKe express better fairness than RED. CHOKe has 70.9% fairness and gCHOKe 87.9%. Fairness of CHOKeD, under single UDP, is 96.6%. Thus, in terms of of fairness, CHOKeD outperforms CHOKe, gCHOKe and RED. Concluding, that CHOKeD link's fairness improves up to almost 8% as compared with gCHOKe, from CHOKe 26% and from RED 43%.

With better fairness, and higher TCP throughput CHOKeD outperforms also in terms of queuing delay as compared with RED, CHOKe and gCHOKe. From Table II, RED has highest queuing delay, *i.e.,* 0.36 seconds and CHOKeD with 0.23 seconds that is lowest. While CHOKe and gCHOKe are close to each other with 0.29 and 0.27 respectively. Consequently, CHOKeD shows better performance in terms of queuing delay too.

### 4.2.5 QUEUE STABILITY

Traffic model 1 from Table I is followed in order to evaluate CHOKeD for the queue stability and simulation result of the average queue size plot have been presented Fig 4.7.

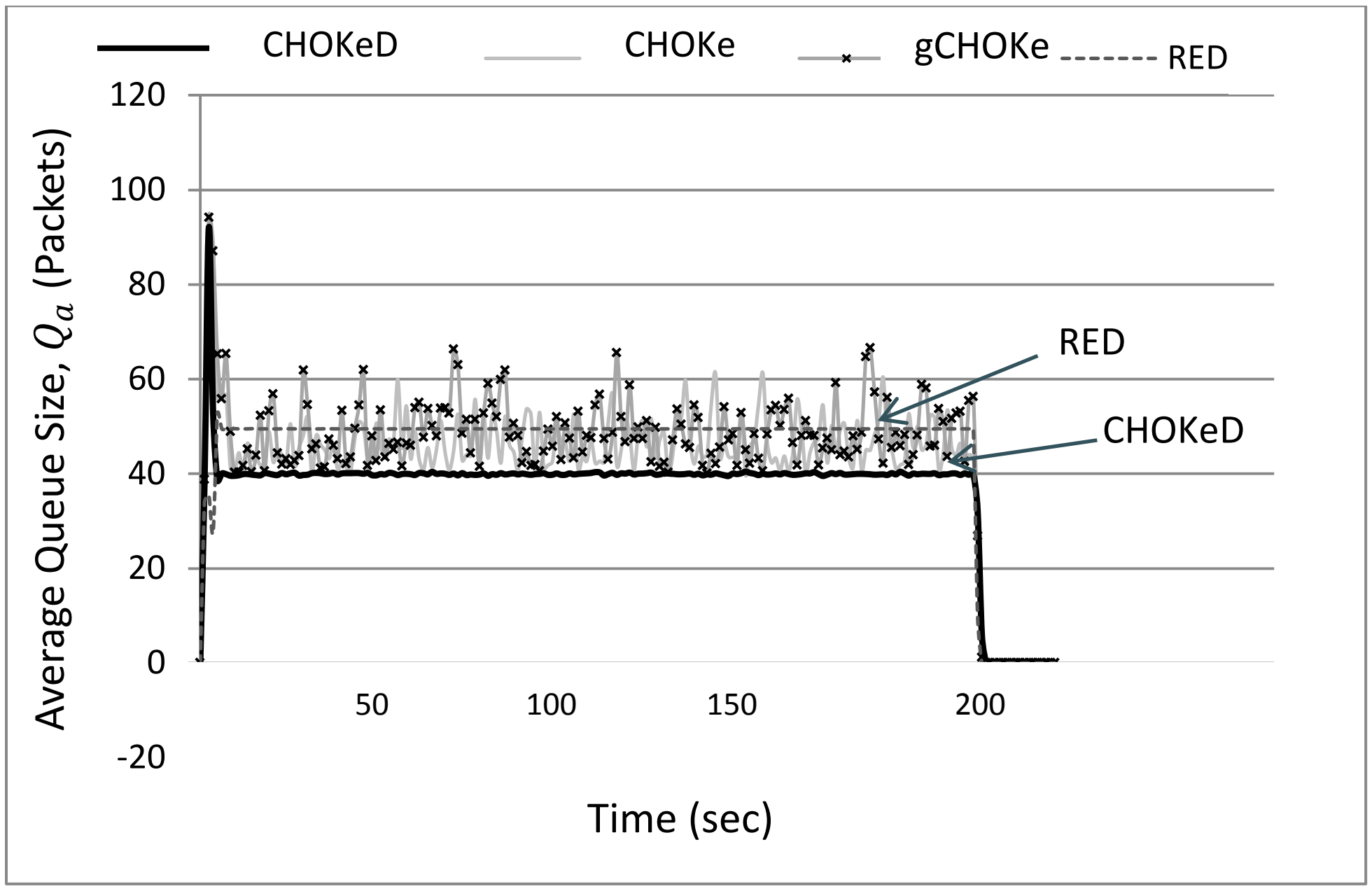

FIG. 4.7: QUEUE STABILITY, UNDER TRAFFIC MODEL 1

The results in Fig 4.7 RED and CHOKeD only display stable behavior, *i.e.*, less oscillation in average queue size $Q_a$. While under CHOKe and gCHOKe $Q_a$, faces high oscillations, *i.e.*, less stable behavior of these two queue models. Result shows another good feature of CHOKeD that it is maintaining average queue size $Q_a$ in stable and control manner. This average queue of CHOKeD is nearly same as the average queue behavior of RED. Thus, CHOKeD show better queue stability than CHOKe and gCHOKe, around the average 40 number of packets with buffer size 100 packets.

## 4.3 Multiple Responsive and Multiple Unresponsive flows

In second simulation phase, multiple UDPs and multiple TCPs are considered. UDPs are taken as 12% of internet traffic as justified in (Abbas *et al.*, 2010). All UDPs have 2 Mb/s rate, twice as compared with link's capacity. Numbers of flows in this simulation are explained in the Table III. For all these flows (1) TCP throughput (2) TCP goodput and (3) fairness, are analyzed, under RED, CHOKe, gCHOKe and CHOKeD.

TABLE III. TRAFFIC MODEL 2

| **TRAFFIC MODEL 2** | | |
|---|---|---|
| Total number of flows | TCP Flows | UDP Flows |
| 25 | 22 | 3 |
| 50 | 44 | 6 |
| 73 | 66 | 9 |
| 100 | 88 | 12 |

### 4.3.1 TCP Throughput Analysis

Traffic model 1 is followed to evaluate the performance of CHOKeD. Fig. 4.8 shows the throughput plot of TCP flows.

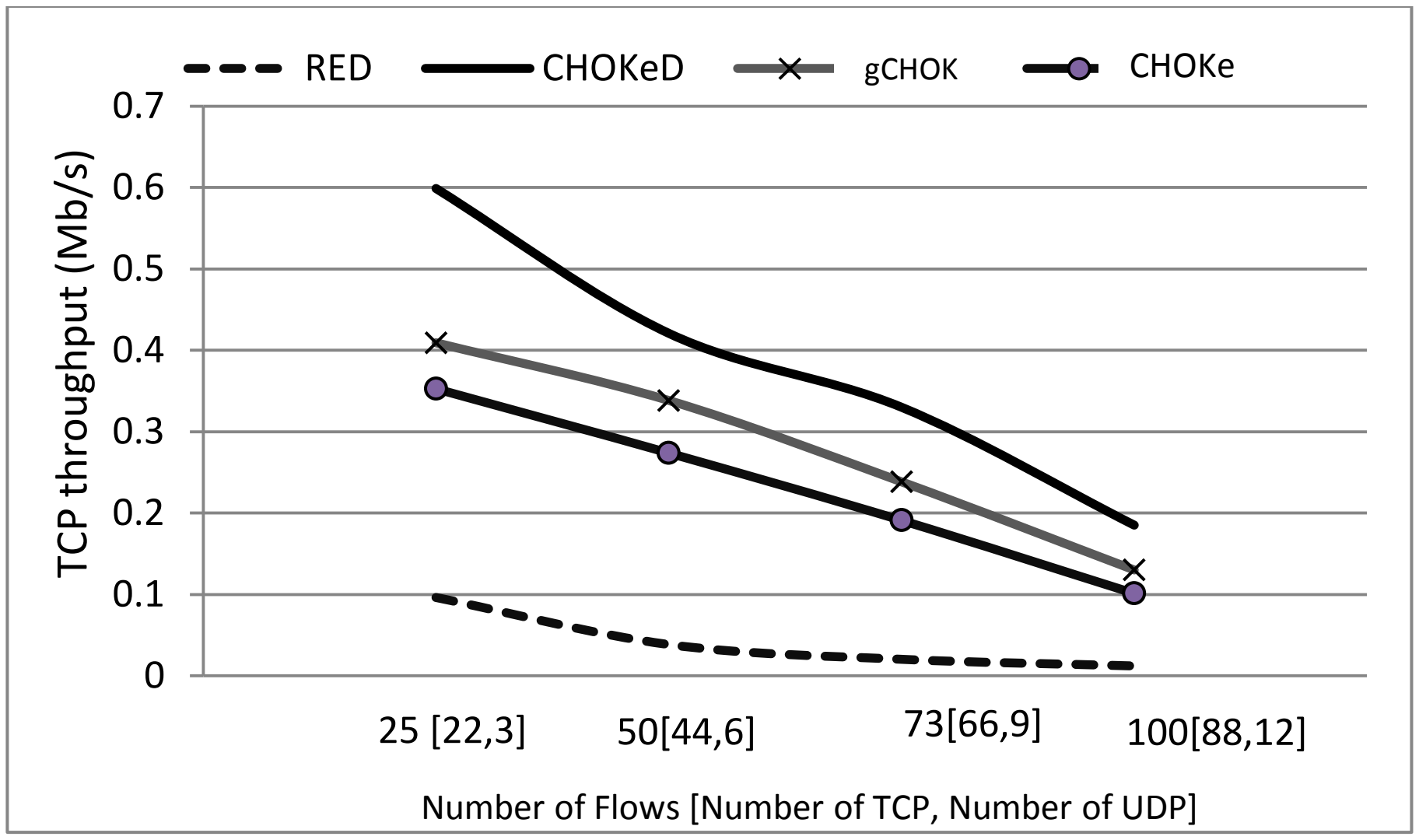

FIG. 4.8: TCP THROUGHPUT OF RED CHOKE, GCHOKE AND CHOKED UNDER TRAFFIC MODEL 2

The graph in Fig 4.8 shows results of four simulations for throughput analysis. CHOKeD is at all above, then gCHOKe , CHOKe and lately RED. In the presence of 3 UDP flows, from 25 flows, CHOKeD yields 0.6 Mb/s TCP throughput while gCHOKe 0.42 Mb/s, CHOKE 0.368 Mb/s while RED with 0.1092 Mb/s. CHOKeD successfully gives better TCP throughput in the presence of 3 unresponsive flow as compared with CHOKe and gCHOKe. CHOKeD behavior is friendly towards responsive traffic. Although CHOKeD punishes unresponsive flows and as a result it gives more room to TCP flows. Moving further and extending this simulation with 50 flows, 44 TCPs and 6 UDPs. During this simulation, TCP is still at higher throughput level under CHOKeD than that of TCPs under gCHOKe, CHOKe and RED. TCP throughputs as: under CHOKeD, TCP attains 0.44 Mb/s, while under gCHOKe 0.33 Mb/s, CHOKe 0.27 Mb/s and with RED 0.05 Mb/s. CHOKeD is friendly towards TCPs even when the number of UDP flows increased up to 6. RED is near the cutoff for TCP as approaching towards zero, just in the presence of 6 UDPs. CHOKeD manages TCP protection better than gCHOKe and CHOKe, clearly visible from Fig 4.7. Furthermore, numbers of flows are increased from 50 to 73 flows with 9 UDPs and 66 TCPs. Under these 73 flows, TCP throughputs are as: CHOKeD 0.325 Mb/s, gCHOKe 0.247 Mb/s, CHOKe 0.192 Mb/s and RED 0.022 Mb/s. CHOKeD TCP flows even get enough good reasonable throughputs. RED getting worse, CHOKe and gCHOKe are better, but still lower than CHOKeD. Even more this simulation continues up to 100 flows with 12 unresponsive flows and 88 responsive. In the scenario RED is completely cut off. CHOKe and gCHOKe are almost at 0.1 Mb/s while CHOKeD nearly at 0.2 Mb/s.

Consequently, although with the increase of unresponsive flows CHOKeD loses its performance, but still CHOKeD provides better protection of TCP flows as compared with gCHOKe, CHOKe and RED. Even in the presence of 12 UDPs CHOKeD manages to provide good buffer space to TCPs rather than completely invaded by unresponsive flows.

### 4.3.2 TCP GOODPUT ANALYSIS

TCP goodput is calculated using Eq. 4.1, for the number of TCPs that varies from 22 to 88 in the presence of unresponsive UDP flows. TCP goodput graph is shown in Fig. 4.9. CHOKeD ahead from the other AQM schemes with 0.063 Mb/s throughputs in the presence of three UDPs while gCHOKe and CHOKe are at 0.038 and 0.033 respectively. RED is near cutoff line. CHOKeD allows TCP flows better bandwidth allocation are compared with gCHOKe, CHOKe and RED. So CHOKeD is friendly towards responsive flows.

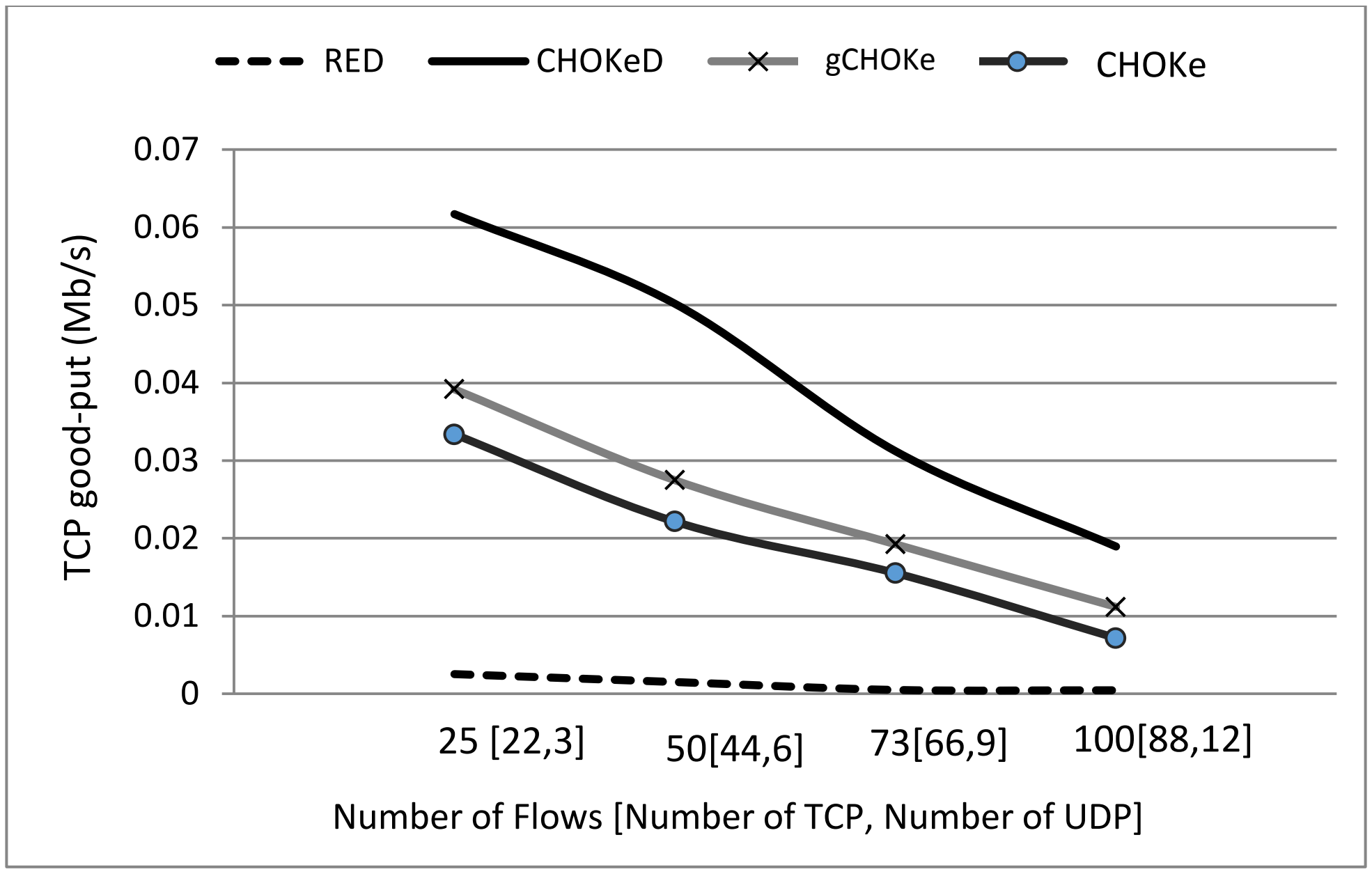

FIG. 4.9: TCP GOODPUT OF RED, CHOKE, GCHOKE AND CHOKED

In the presence of CHOKeD, TCP flows are able to attain their bandwidth share. Further, this simulation is extended for 50 flows, including 44 TCPs and 6 UDPs. In this scenario, CHOKeD is also at a higher level of TCP goodput than gCHOKe, CHOKe and RED. In the next scenario, the numbers of flows are increased from 50 to 73 with 66 TCPs and 9 UDPs. Even in the presence of 9 UDPs, TCP flows still survive and get 0.03 Mb/s goodput. At last, the numbers of flows are increased from 73 to 100 and at this level CHOKeD is still far better than CHOKe and gCHOKe, while RED is completely shut off.

Thus, TCPs flows get better goodput values under CHOKeD even under 12 UDPs and still getting bandwidth share. Failure of RED is completely visible as in the presence of 9 UDPs, it is completely shut down, no more buffer space for TCPs.

### 4.3.3 FAIRNESS

CHOKeD fairness is observed to evaluate more extensively, in the presence of multiple TCPs and multiple UDPs. Fig. 4.9 shows a fairness plot, in which the numbers of flows are varied from 25 to 100. In the presence of 22 TCPs and 3 UDPs, CHOKeD fairness is 0.88. While gCHOKe with 0.80, CHOKe 0.68 and RED at 0.33. CHOKeD is fairer than all others.  Similarly, in the presence of 100 flows CHOKeD shows 0.42, while gCHOKe 0.36 and CHOKe 0.35. RED shows 0.187 fairness index value.

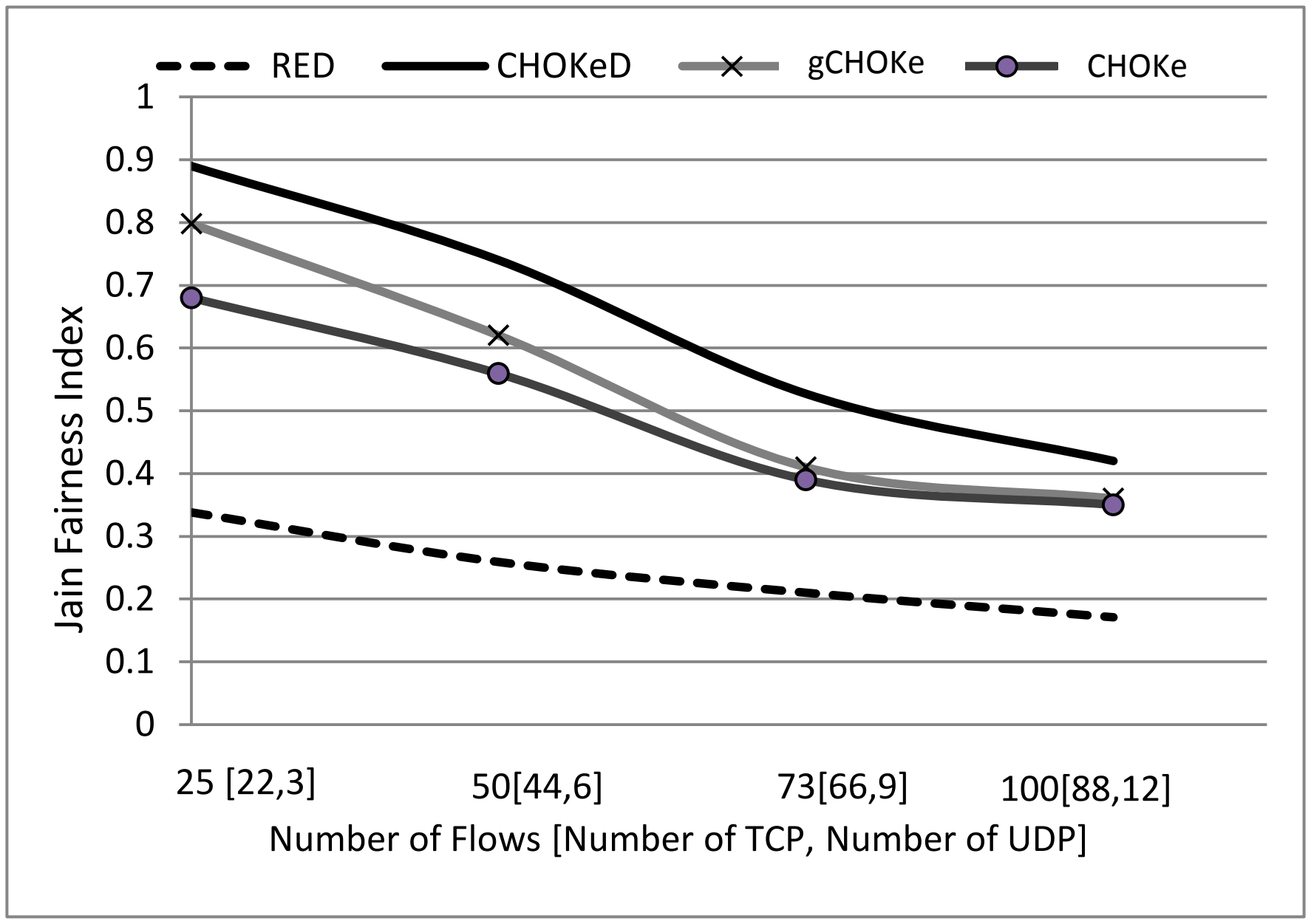

FIG. 4.10: FAIRNESS OF RED, CHOKE, GCHOKE AND CHOKED UNDER TRAFFIC MODEL 2

While summarizing all above, CHOKeD confirms itself as better in terms of fairness as compared with other three AQM schemes. Its fairness is due to good punishment criteria for unresponsive flows and better bandwidth allocation for responsive flows. As UDPs throughput is minimized by CHOKeD so TCPs have more chance to attain their buffer shares. Eventually, all flows attain almost their ideal fair share.

### 4.3.4 QUEUING DELAY ANALYSIS

Queuing delay is computed by varying the number of flows from 25 to 100, from the Table II, and plot is shown in Fig. 4.11.

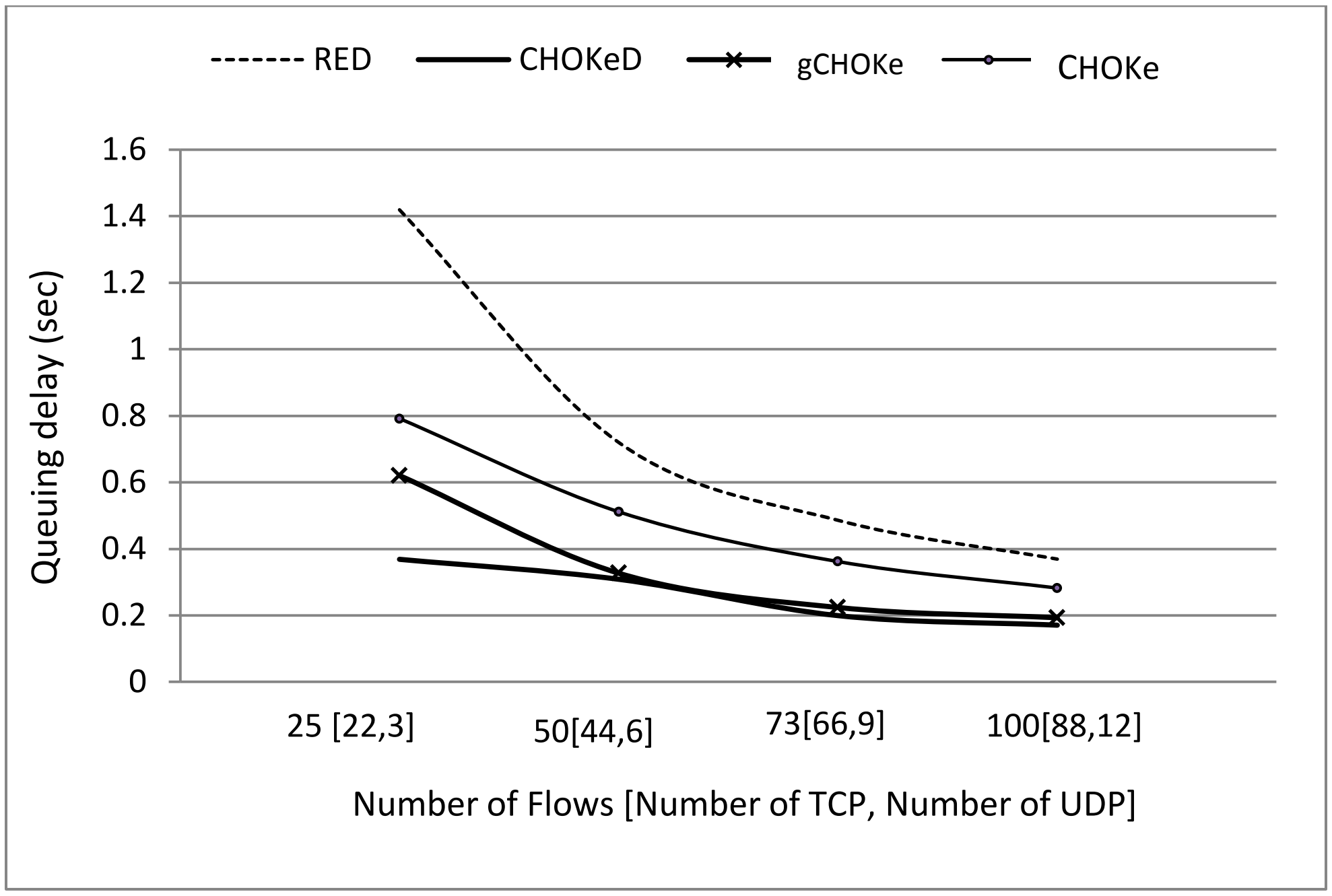

FIG. 4.11: QUEUING DELAY UNDER TRAFFIC MODEL 2

From Fig 4.11, RED has maximum queuing delay and CHOKeD at the lowest level, while CHOKe and gCHOKe are in between for all these four simulations. Under the CHOKeD queuing delay is lowermost as more number of TCP packets allowed to reside and delivering towards the destination. Protection against UDPs lead CHOKeD towards low queuing delay.

## 4.4 INTER-TCP PROTOCOL FAIRNESS

Sometimes it is realized that AQM scheme may be biased with some versions of TCP and shows good performance with only those versions of TCP (Bhatti *et al.,* 2008). This argument is considered and justify with the results, while taking simulation of 1 UDP, 1 TCP Reno and 1 TCP Vegas. Simulation result is shown in TABLE IV.

TABLE IV. THROUGHPUT UNDER TCP RENO AND VEGAS

| AQM | TCP Reno Throughput (kb/s) | TCP Vegas Throughput (kb/s) | UDP Throughput (kb/s) | SDV (kb/s) |
|---|---|---|---|---|
| CHOKeD | 220.1 | 211.5 | 390.3 | 100.84 |
| gCHOKe | 209.4 | 192.1 | 424.6 | 129.53 |
| CHOKe | 207.9 | 196.3 | 453.1 | 145.03 |
| RED | 110.4 | 98.4 | 574.5 | 271.48 |

TABLE IV depicts throughputs attained by TCP Reno and TCP Vegas under CHOKeD, CHOKe and gCHOKe. Table III shows, lowest standard deviation value, *i.e.,* 100.84 Kb/s in case of CHOKeD as TCP Vegas has 211.5 Kb/s and Reno has 220.1 Kb/s throughputs. Graph results support that CHOKeD is not biased for TCP Vegas and TCP Reno.

## 4.5 FAIRNESS UNDER TCP WITH DIFFERENT RTTS

TCPs with smaller RTTs send more often acknowledgment packets than the TCPs with larger RTTs, these short RTT TCPs successfully attain more bandwidth space and affects TCPs with large RTTs (Chatranon *et al.,* 2004). Simulation is carried out to evaluate CHOKeD with different round trip times (RTTs), TCP flows. Traffic model 2 is used to evaluate, in which half of TCPs have 50ms RTT, while others have 20 ms. Simulation is carried out for 200 seconds with a buffer size of 100 packets. Simulation plot is shown in Fig. 4.12, fairness of RED, CHOKe, gCHOKe and CHOKeD. CHOKeD losses some fairness under this scenario, but still able to outperform than RED, CHOKe and gCHOKe. Surprisingly RED shows almost linear behavior under different RTT based TCPs.

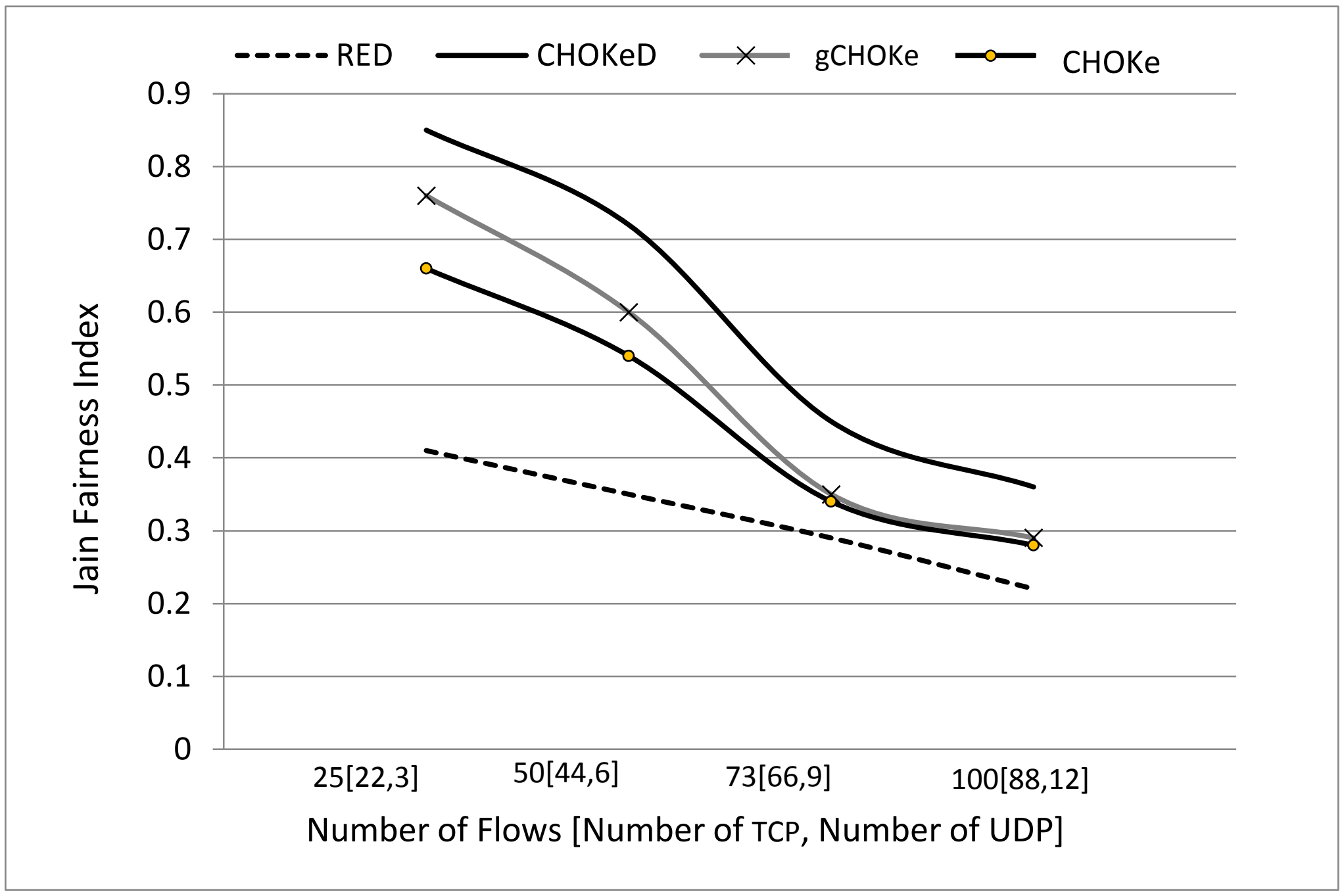

FIG. 4.12: FAIRNESS WITH DIFFERENT RTTS TCP UNDER TRAFFIC MODEL 2

CHOKe and gCHOKe also lose their previous fairness level. Actually TCPs with smaller RTTs send more often acknowledgment packets than with larger RTTs, these TCP successfully attain more bandwidth space and affects TCPs with large RTTs. That is why RED, CHOKe, gCHOKe and CHOKeD losses their fairness under TCPs with different

RTTs. From Fig. 4.12 CHOKeD still fights against these unfair TCPs and show higher fairness as compared with RED, CHOKe and gCHOKe.

## 4.6 DIFFERENT BUFFER SIZE

As there is a tradeoff between buffer size and link delay, researchers suggest router of small buffer size (Adam, 2013). That is why in market different routers with different buffer sizes are available. To discuss robustness of CHOKeD, simulations for different buffer sizes of router, are carried out. Simulations with a buffer of size 300 and 500 packets are performed. Comparisons are shown in TABLE V and TABLE VI.

Simulations consider 33 TCP and 1 UDP flow to calculate TCP throughput, UDP throughput, TCP goodput, queuing delay and fairness. From Table IV CHOKeD shows 95% fairness and 197 seconds queuing delay.

TABLE V. SIMULATION RESULTS WITH BUFFER OF SIZE 300 PACKETS

| AQM | TCP throughput (Mbs) | UDP throughput (Mbs) | TCP Goodput (Mbs) | Queuing delay (ms) | Fairness |
|---|---|---|---|---|---|
| RED | 0.110883 | 0.530768 | 0.07176 | 1639.33 | 0.49 |
| CHOKeD | 0.615168 | 0.204883 | 0.43796 | 197.35 | 0.95 |
| CHOKe | 0.531019 | 0.255005 | 0.386962 | 292.547 | 0.75 |
| gCHOKe | 0.567731 | 0.228966 | 0.421 | 322.049 | 0.83 |

TABLE VI. SIMULATION RESULTS WITH BUFFER OF SIZE 500 PACKETS

| AQM | TCP throughput (Mbs) | UDP throughput (Mbs) | TCP Goodput (Mbs) | Queuing delay (ms) | Fairness |
|---|---|---|---|---|---|
| RED | 0.089986 | 0.542372 | 0.05794 | 3138.37 | 0.38 |
| CHOKeD | 0.574147 | 0.213571 | 0.434162 | 401.43 | 0.89 |
| CHOKe | 0.521647 | 0.257291 | 0.390724 | 436.898 | 0.72 |
| gCHOKe | 0.543582 | 0.243744 | 0.407763 | 417.76 | 0.79 |

With 500 packet size buffer simulations carried out and results shown in TABEL VI. As the buffer size increases, so more buffer space available for unresponsive flows, that is why CHOKeD, CHOKe and gCHOKe loses their fairness. But CHOKeD is 89% fair with 401.43 seconds queuing delay. Under CHOKeD TCP throughput and goodput is higher, because of low UDP throughput.

## 4.7 WEB-MIXED TRAFFIC

Web traffic contains long-lived TCPs and short-lived TCPs. AQM schemes fail to ensure fairness in the presence of many long-lived TCP flows. Using, 15 FTP and 15 HTTP based TCP connections with 1 CBR based UDP, web-mixed experiment is performed using dumb-bell topology and plot is described in Fig. 4.13. The proposed scheme outperforms and protects more short-lived TCP in the presence of long-lived TCPs.

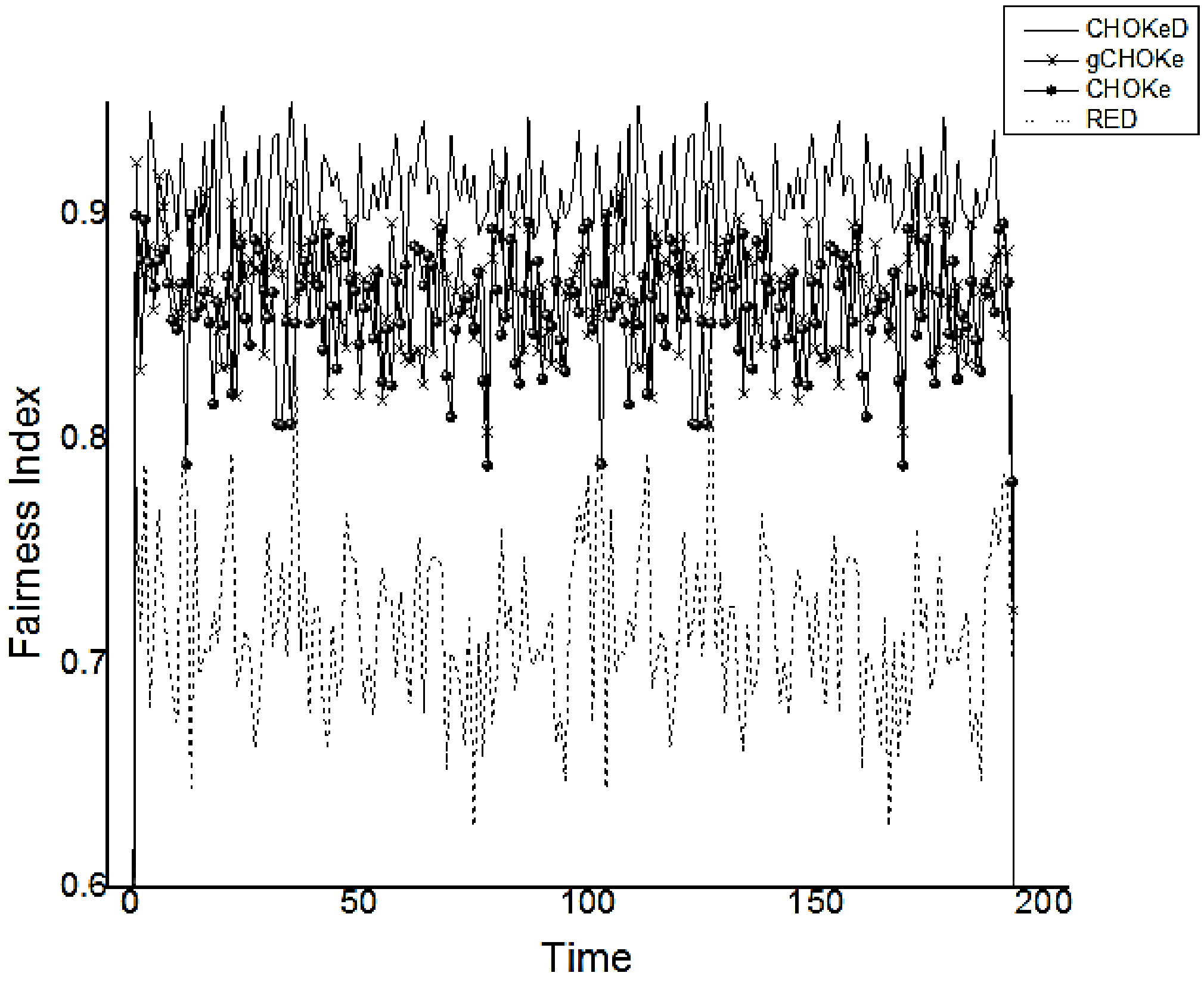

FIG. 4.13: FAIRNESS INDEX UNDER WEB MIX TRAFFIC

Fairness plot is shown in Fig. 4.13 verses time. CHOKeD ensures more fairness with at the top with a highest fairness index value. CHOKeD not only normalize the UDP flow, but also successfully protect shot-lived TCPs. RED, gCHOKe and CHOKe have low fairness index values as compared with CHOKe. CHOKeD drawing factor supports TCPs and penalizes UDP flows more aggressively. Thus, CHOKeD behaves well under web-mixed traffic also.

CHOKeD gives good results with 92% fairness among RED, CHOKe and gCHOKe but it protects only up to some level HTTP based TCP traffic, while RED, CHOKE and gCHOKe's performance is much better from all previous comparisons.

## 4.8 Time Complexity of CHOKeD

CHOKeD is completely stateless approach and its complexity is $O(D)$, where $D = D_r + D_f$ is the complete drawing factor. CHOKeD is slightly more complex than RED and CHOKe as its drawing factor is higher than CHOKe to punish misbehaving traffic more rigorously. This $O(D)$ complexity is slightly higher than CHOKe's complexity which is $O(1)$ but much lesser than the complexity of per-flow maintaining AQM schemes like RED-PD (Mahajan and Floyd, 2001), which has $O(n)$ complexity. CHOKeD complexity is independent of number of flows.

# CHAPTER 5

# 5 CONCLUSIONS AND FUTURE WORK

## 5.1 CONCLUSIONS

The basic design purpose of CHOKeD is an active queue management based system that certifies high level fairness among all traffic flows, with less complexity. The CHOKeD is a stateless and simple match-drop based approach for fairness implementation over network routers. CHOKeD follows the design requirements as well as improvements over the traditional AQM protocols. CHOKeD also focuses on less queuing delays and is compatible with small buffer size to remove extra-transmission overheads. CHOKeD is designed to enforce fairness and to protect responsive traffic.

CHOKeD successfully shields responsive TCP traffic and penalizes misbehaving UDP traffic to promise a good level of fairness. Through a series of simulations, it is validated that CHOKeD is a better AQM approach for fairness. CHOKeD fairness mechanism is depending upon its drawing factor. Location based and dynamic numbers of the candidate packet selection in match drop comparisons, helps to control aggressiveness of unresponsive traffic. CHOKeD gives more protection to TCP-friendly traffic as compared with many traditional and recent AQM schemes like RED, CHOKe and gCHOKe. CHOKeD also outperforms RED, CHOKe and gCHOKe, in terms of fairness, high throughput and low queuing delay under single and many unresponsive flows.

## 5.2 Future Work

For future guidelines, CHOKeD analytical model may be studied to strengthen and identify weaknesses. Simulations and experimental work might be performed on a network router to analyze CHOKeD behavior under real time network traffic.

# APPENDIX A
# CHOKeD ALGORITHM

CHOKeD algorithm is described Algorithm I, and is given below:

**ALGORITHM I. CHOKE DESCENDANT (CHOKED)**

{Initialization}
$D_i \leftarrow 0$
$D_r \leftarrow 0$
$D_f \leftarrow 0$
$S \leftarrow 0$
$r_{rare} \leftarrow 0$
$r_{front} \leftarrow 0$
$\{\boldsymbol{packet\ admission\ action}\}$
**if** $Q_a < T_{min}$
  admit $p$
**else if** $Q_a > T_{max}$
  drop $pk$
**else**
  $S \leftarrow \frac{Q_c}{2}$
  $r_{rare} \leftarrow Q_c - S$
  $r_{front} \leftarrow \| 0 - r_{rare} \|$
  $D_r \leftarrow round\left(\frac{(Q_c * \sqrt{B})}{(T_{max} - T_{min}) \quad * \ln(B)}\right)$
  $D_f \leftarrow round(\frac{D_r}{2})$
$\{\boldsymbol{packet\ drawing\ action\ for\ rare\ region}\}$
**while** $D_r > 0$ **do**
  $D_r \leftarrow D_r - 1$
  $Draw\ packet\ pk'\ from\ the\ buffer\ randomly$
  **if** $ID(pk') = ID(pk)$
    drop $pk'\ and\ pk$
  **end if**
**end while**
$\{\boldsymbol{packet\ drawing\ action\ for\ front\ region}\}$
**while** $D_f > 0$ **do**
  $D_f \leftarrow D_f - 1$
  $Draw\ packet\ pk'\ from\ the\ buffer\ randomly$
  **if** $ID(pk') = ID(pk)$
  drop $pk'\ and\ pk$
  **end if**
**end while**
**else**

```
32.       admit pk
33.    end if
```

**Input Parameters:**

$Q_a$: Average queue size
$Q_c$: Current queue size
$T_{min}$: Average queue size minimum threshold
$T_{max}$: Average queue size maximum threshold
$D_i$: Drawing factor
$D_r$: Drawing factor for rear region
$D_f$: Drawing factor for front region
S: midpoint of the current queue
$B$ : Buffer size
$r_{rare}$: rare end of queue
$r_{front}$: front end of queue
*pk':* packet from the queue
*pk: a*rrived packet

Algorithm description is given as: Firstly, from line 1 to 6, input variables are initialized, if average queue size $Q_a$ is less than the average queue size minimum threshold $T_{min}$ CHOKeD will admit newly arrived packet $pk$. From line 7 to 26, If $Q_a$ is less than average queue size maximum threshold $T_{max}$ but greater than $T_{min}$, CHOKeD obtains midpoint of current queue size $Q_c$, and selects two queue regions front queue region $r_{front}$ and rear queue region $r_{rear}$. Now it calculates drawing factor $D_r$ and draws $D_r$ packets from $r_rear$ and compare it with $pk$. If $pk'$ (packets from queue) and $pk$ have same flow id, both deleted. From line 27 to 34, as no packet from queue has same id as of the $pk$, now CHOKeD draws $D_f$ packets and compare it with $pk$. If $pk'$ and $pk$ have same flow id, both are dropped. From line 35 to 38, as $pk$ does not match with the flow id of any packet from the queue, so it will be admitted.

# APPENDIX B

## NS2 INSTALLATION OVER FEDORA 14

Download the all-in-one package

```
$ wget
http://downloads.sourceforge.net/project/nsnam/allinone/ns-
allinone-2.34/ns-allinone-2.34.tar.gz
```

Un-compress the package

```
$ tar xf ns-allinone-2.34.tar.gz
```

Install needed pacakges

```
# yum install gcc make libX11-devel libXt-devel libXmu-devel
```

Ns-2 requires older version of gcc. So install gcc-34 and gcc-34-c++ for it

```
# yum install compat-gcc-34 compat-gcc-34-c++
```

Configure the environmental variables for ns-2 and nam, and add the executable to the PATH so that user can use ns and nam directly.

Add to *~/.bashrc* if you use bash

```
NS_HOME=/full/path/to/ns-allinone-2.34

PATH=$NS_HOME/bin:$NS_HOME/tcl8.4.18/unix:$NS_HOME/tk8.4.18/unix
:$PATH

export PATH
```

# APPENDIX C

## HOW TO WRITE NEW AQM PROTOCOL

**0. Download ns2, Compile it and Install it.**

This is a preparation step. It's not the focus of this post, please refer to reference 1 for details. Note that since we're going to modify NS2 by adding our own protocol to it, we'll need to download and source code and build it ourselves.

**1. Add C++ Source Files and Update Makefile**

For AQM protocols, their source files are placed under the folder ns-allinone-2.35/ns-2.35/queue/. Suppose our new protocol is named MOD, and it's implemented in mod.cc and mod.h source files. Simply copy mod.cc and mod.h files to ns-2.35/queue/ folder.

Next we'll need to modify the makefile so that mod.cc and mod.h can be compiled. Open ns-allinone-2.35/ns-2.35/Makefile. Look for queue/red.o and add queue/mod.o. So the line will become,

*adc/simple-intserv-sched.o queue/red.o queue/mod.o *

Save the changes and type "make" at ns-allinline-2.35/ns-2.35/ directory. The newly added protocol will be built.

**2. Add MOD for TCL Scripting**

NS2 uses TCL for simulation configuration. So far we only builds the protocol in C++, there're still some work needs to be done in order for the newly added protocol to be used in TCL.

**2.1 Changes Files under ns-2.35/tcl/lib/**

There're a few files needs to change in order for MOD for work. I don't want this post to be lengthy, so it's not shown in detail, you can download the file at the end of the post, and search for MOD to see the places that have been changed.

The list of files need to change including ns-queue.tcl, ns-default.tcl, ns-route.tcl, ns-compat.tcl, and ns-lib.tcl.

### 2.2 Compile Again

Go to directory ns-2.35 in command line, and type “su” then “make” to build ns2.

# C++ CODE COMPILATION

- Type “su” (enter into root)
- Go to ns-allinline-2.35/ns-2.35/ directory then type “make clean”
- Type “./configure”
- Finally type “make”

Or simply in ns-allinline-2.35/ns-2.35/ directory type, “make clean; ./configure; make”